\documentclass[aps,prd,showpacs,showkeys]{revtex4-1}
\usepackage{amsmath}
\usepackage{amssymb}
\usepackage{placeins}

\usepackage{graphicx}

\begin{document}

\title{Outlook on the Higgs particles, masses and physical bounds in
  the Two Higgs-Doublet Model} 

\author{S.R. Ju\'{a}rez W.}

\thanks{Proyecto SIP: 20110368, Secretar\'{\i}a de Investigaci\'on y
  Posgrado, Beca EDI y Comisi\'on de Operaci\'on y Fomento de
  Actividades Acad\'emicas (COFAA) del IPN}

\email{rebeca@esfm.ipn.mx} 

\author{D. Morales C.}  

\thanks{Becario del Consejo Nacional de Ciencia y Tecnolog\'{\i}a
  (CONACyT), Mexico}

\email{Damoralesc@hotmail.com} 

\affiliation{Departamento de F\'{\i}sica, Escuela Superior de
  F\'{\i}sica y Matem\'{a}ticas, Instituto Polit\'{e}cnico Nacional,
  U.P. ``Adolfo L\'{o}pez Mateos'' C.P.~07738, M\'{e}xico D.F.}

\author{P. Kielanowski}

\affiliation{Departamento de F\'{\i}sica,
  Centro de Investigaci\'{o}n y Estudios Avanzados\\
  Av. IPN 2508, C.P. 07360, M\'exico, D.F.}  

\thanks{Work was partly supported by Polish Ministry of Science and
  Higher Education Grant N~N202~230337} \email{kiel@fis.cinvestav.mx}

\begin{abstract}
  The Higgs sector of models beyond the standard model requires
  special attention and study, since through them, a natural
  explanation can be offered to current questions such as the big
  differences in the values of the masses of the quarks (hierarchy of
  masses), the possible generation of flavor changing neutral currents
  (inspired by the evidence about the oscillations of neutrinos),
  besides the possibility that some models, with more complicated
  symmetries than those of the standard model, have a non standard low
  energy limit. The simplest extension of the standard model known as
  the two-Higgs-doublet-model (2HDM) involves a second Higgs
  doublet. The 2HDM predicts the existence of five scalar particles:
  three neutral ($A^{0}$), ($h^{0}$, $H^{0}$) and two charged
  ($H^{\pm}$). The purpose of this work is to determine in a natural
  and easy way the mass eigenstates and masses of these five
  particles, in terms of the parameters $\lambda _{i}$ introduced in
  the minimal extended Higgs sector potential that preserves the CP
  symmetry.  We discuss several cases of Higgs mixings and the one in
  which two neutral states are degenerate.  As the values of the
  quartic interactions between the scalar doublets are not
  theoretically determined, it is of great interest to explore and
  constrain their values, therefore we analize the stability and
  triviality bounds using the Lagrange multipliers method and
  numerically solving the renormalization group equations. Through the
  former results one can establish the region of validity of the model
  under several circumstances considered in the literature.
\end{abstract}


\maketitle
\section{Introduction\label{sec1}}

The Standard Model (SM) in high energy physics~\cite{wsg} has been
remarkably successful in: describing the properties of elementary
particles, predicting the existence of the quarks $c$, $t$ and $b$,
and the third generation of leptons $\tau $- $\nu _{\tau }$, the
existence of the eight gluons, and the weak bosons $W^{\pm }$, $Z^{0}$
before their discovery, predicting parity violating
neutral-weak-currents, and in being consistent with all the
experimental results~\cite{PDG,CERN}. However, the SM falls short of
being a complete theory of the fundamental interactions because of its
lack of explanation of the probable unification of the fundamental
interactions, the pattern and disparity of the particle masses (mass
hierarchy), the origin of the CP violation in nature, the
matter-antimatter asymmetry, the pattern of quark mixing, lepton
mixing and the reason why there are 3~generations.

As a partial solution to confront these deficiencies, a large number
of parameters must be put in ``by hand'' into the theory (rather than
being derived from first principles), such as the three gauge
couplings ($g_{1}$, $g_{2}$ and $g_{3})$, nine fermionic masses (six
quarks and three leptons), the Weinberg angle ($\theta _{w}$), four
quark-mixing parameters (CKM) and two more parameters in relation to
the Higgs potential ($\mu $ and $\lambda $).

One of the most subtle aspects of the model is associated with the Higgs
sector~\cite{Higgs}. The Higgs field and its non-vanishing vacuum
expectation value (vev) is the essential ingredient to carry out the
spontaneous symmetry breaking (SSB) required to transform the hypothetical
massless particles in the Lagrangian into the actual massive physical
particles. However, the Higgs particle has not yet been discovered.

In this paper we study the extension of the SM with two Higgs doublets
(2HDM) that presents the challenge that the quartic interactions
between the scalar doublets are not theoretically determined.  This
model is studied mainly for three reasons. The first one is that the
2HDM has a much richer Higgs spectrum (3 neutral and 2 charged
Higgses) and a different high energy behavior. This makes that a lower
mass than in the SM Higgs is permitted. Another reason may be that a
different pattern of hierarchy of the Yukawa couplings is possible,
because of the presence of two independent vacuum expectation values
of the Higgs fields~\footnote{The importance of such analysis can be
  seen for example in the Higgs search scenarios, e.g., David
  L\'opez-Val and Joan Sol\`a, Phys. Rev.~D \textbf{}81, (2010), Joan
  Sol\`a, David L\'opez-Val, arXiv:1107.1305 [hep-ph].}.  The third
reason is that the Higgs sector of the Minimal Supersymmetric Standard
Model (MSSM) contains two Higgs doublets, so the Higgs sectors of the
MSSM and the 2HDM are similar and the study of the 2HDM model may give
important information on the properties of the Higgs sector in the
MSSM~\footnote{See for example Apostolos Pilaftsis and Carlos
  E.M. Wagner, Nucl. Phys.~\textbf{553}, 3 (1999).}.

In Section~\ref{sec2} we introduce the potential for the 2HDM in a
special parametrization, and briefly discuss the SSB. In
Sections~\ref{sec3} and~\ref{sec4} we present the Higgs mass matrix
and its diagonalization method, the mass spectrum, mass eigestates,
and special cases of mixing, where the Higgs masses are simply related
to the parameters of the potential. In Section~\ref{sec5} we obtain
and classify the constrains for the quartic couplings derived from the
mass formulas, from the vacuum stability principle through the
Lagrange multipliers method, and by imposing extreme stability
conditions. In Section~VI we numerically solve the set of the
renormalization group equations from which through triviality
principle the physical bounds of the model are determined under
different conditions. Finally, Section~\ref{sec7} is devoted to the
presentation of the results and the conclusions.

\section{The two-Higgs doublet model}\label{sec2}

In the SM the fermion masses arise, after the SSB, from the couplings
between the fermions and a single Higgs doublet. The mass ratio of the
$b$ and $t$ quark is of the order of $1/40.$ To understand in a
natural way the origin of this difference in the values of the masses
of the third generation of quarks, one can assume the existence of a
second Higgs-doublet in the Higgs sector of the SM.  In this context
one assumes that the quark $t$ obtains its mass through the $\Phi
_{1}$ doublet and the quark $b$ from another doublet $\Phi
_{2}$~\footnote{There are also other scenarios for the quark mass
  generation in the 2DHM but we will not be considering them
  here.}. In this way one can explain in a more natural way the
hierarchy problem of the Yukawa couplings, as long as the free
parameters of the new model acquire the appropriate values.

The Higgs sector of the 2HDM consists of two identical
(hypercharge-one) scalar doublets $\Phi _{1}$ and $\Phi _{2}$. There
are several proposals for the Higgs potential to describe the physical
reality in the framework of the 2HDM~\cite{Inoue,GHKD}. The potential
we consider in this paper is compatible with Ref.~\cite{Kominis}. It
is such that the CP symmetry (charge-conjugation and parity) is
conserved, the neutral-Higgs-mediated flavor-changing neutral currents
(FCNC) are suppressed in the leptonic sector, and in the quark-sector
they are also forbidden by the GIM mechanism~\cite{GIM} in the one
loop approximation. In the Lagrangian $\mathcal{L}$ in which we leave
out the leptonic terms,
\begin{equation}
  \mathcal{L=L}_{gf}+\mathcal{L}_{Kin}+\mathcal{L}_{Y}-V
\end{equation}
the $\mathcal{L}_{gf}$ and $\mathcal{L}_{Kin}$ correspond to kinetic
parts of quarks and bosons and they contain the covariant derivatives
that provide the interactions among the gauge bosons and the Higgs
bosons.  They also give rise, after the SSB, to the masses of the
gauge bosons (mediators of the electroweak interactions). The fermion
masses are generated from the Yukawa couplings in $\mathcal{L}_{Y}$
\begin{equation}
\mathcal{L}_{Y}=\sum_{i.j}\left( g_{ij}^{(u)}\overline{\psi }_{Li}\Phi
_{1}^{c}u_{Rj}+g_{ij}^{(d)}\overline{\psi }_{Li}\Phi _{2}d_{Rj}\right) ,
\end{equation}
between the Higgs bosons and the quarks. In $\mathcal{L}_{Y}$, the
$g_{ij}^{(u,d)}$ are the Yukawa coupling matrices. The superscripts
$(u,d)$ refer to the up and down sectors of quarks, respectively and
the subscripts $\left( L,R\right) $ correspond to the left handed
doublets and right handed singlets in the quark sector. In this paper,
we will focus our attention on the potential $V$.
\subsection*{The Higgs potential}
 The Higgs potential depends on seven real parameters $\mu
_{1}^{2},\mu _{2}^{2}$ and $\lambda _{i\,}\,\,(i=1...,5)$ from which
the five Higgs masses come up after the SSB. The most general
renormalizable $SU(2)\times U(1)$ invariant Higgs potential, that
preserves a CP and a Z$_{2}$ symmetry ($\Phi _{1}\rightarrow \Phi
_{1},\,\,\Phi _{2}\rightarrow -\Phi _{2})$ is given by
\begin{multline}
  V =\mu _{1}^{2}\Phi _{1}^{\dagger }\Phi _{1}+\mu _{2}^{2}\Phi
  _{2}^{\dagger }\Phi _{2}+\lambda _{1}\left( \Phi _{1}^{\dagger }\Phi
    _{1}\right) ^{2}+\lambda _{2}\left( \Phi _{2}^{\dagger }\Phi
    _{2}\right) ^{2}+\lambda _{3}\left( \Phi _{1}^{\dagger }\Phi
    _{1}\right) \left( \Phi
    _{2}^{\dagger }\Phi _{2}\right)\\
  +\lambda _{4}\left( \Phi _{1}^{\dagger }\Phi _{2}\right) \left( \Phi
    _{2}^{\dagger }\Phi _{1}\right) +\frac{1}{2}\lambda _{5}\left[
    \left( \Phi _{1}^{\dagger }\Phi _{2}\right) ^{2}+\left( \Phi
      _{2}^{\dagger }\Phi _{1}\right) ^{2}\right] .\label{ec1}
\end{multline}
For the sake of simplicity a special basis is introduced
\begin{equation}
  A=\Phi _{1}^{\dagger }\Phi _{1},\quad B=\Phi _{2}^{\dagger }\Phi
  _{2},\quad C^{\prime }=D^{{\prime}{\dagger }}=\Phi _{1}^{\dagger }\Phi _{2}.
\end{equation}
In this basis
\begin{equation}
V=\mu _{1}^{2}A+\mu _{2}^{2}B+\lambda _{1}A^{2}+\lambda _{2}B^{2}+\lambda
_{3}AB+\lambda _{4}C^{\prime }D^{\prime }+\frac{1}{2}\lambda _{5}\left[
C^{\prime 2}+D^{\prime 2}\right] .\label{unoML}
\end{equation}
The two Higgs doublets can be represented by eight real fields $\phi
_{i},\,i=1,\ldots,8$,
\begin{equation}
\Phi _{1}=\left(
\begin{array}{l}
\phi _{1}+i\phi _{2} \\
\phi _{3}+i\phi _{4}
\end{array}
\right) ,\quad \Phi _{2}=\left(
\begin{array}{l}
\phi _{5}+i\phi _{6} \\
\phi _{7}+i\phi _{8}
\end{array}
\right) .\,  \label{geig}
\end{equation}
If charge is conserved and there is no CP violation in the Higgs sector, 
after the SSB, the non-vanishing vacuum expectation values of the fields $\phi _{3}$
and $\phi _{7}$ are real,  
\begin{equation}
\left\langle \phi _{3}\right\rangle =\frac{v_{1}}{\sqrt{2}},\quad
\left\langle \phi _{7}\right\rangle =\frac{v_{2}}{\sqrt{2}},
\end{equation}
\begin{equation}
  \left\langle \phi _{1}\right\rangle =\left\langle \phi _{2}\right\rangle
  =\left\langle \phi _{4}\right\rangle =0,\quad \left\langle \phi
    _{5}\right\rangle =\left\langle \phi _{6}\right\rangle =\left\langle \phi
    _{8}\right\rangle =0.
\end{equation}
In terms of the fields $\phi _{i},$ the hermitian basis is given by
\begin{equation}
  \begin{split}
  A &=\phi _{1}^{2}+\phi _{2}^{2}+\phi _{3}^{2}+\phi _{4}^{2},\quad B=\phi
  _{5}^{2}+\phi _{6}^{2}+\phi _{7}^{2}+\phi _{8}^{2},  \\
  C^{\prime } &=\phi _{1}\phi _{5}+i\phi _{1}\phi _{6}-i\phi _{2}\phi
  _{5}+\phi _{2}\phi _{6}+\phi _{3}\phi _{7}+i\phi _{3}\phi _{8}-i\phi
  _{4}\phi _{7}+\phi _{4}\phi _{8},  \\
  D^{\prime } &=\phi _{1}\phi _{5}+i\phi _{2}\phi _{5}-i\phi _{1}\phi
  _{6}+\phi _{2}\phi _{6}+\phi _{3}\phi _{7}+i\phi _{4}\phi _{7}-i\phi
  _{3}\phi _{8}+\phi _{4}\phi _{8}.
\end{split}
\end{equation}
and after the SSB they become
\begin{equation}
  \left\langle A\right\rangle =\frac{1}{2}v_{1}^{2},\quad \left\langle
    B\right\rangle =\frac{1}{2}v_{2}^{2},\quad \left\langle C^{\prime
    }\right\rangle = \left\langle D^{\prime
    }\right\rangle =\frac{1}{2}v_{1}v_{2}.
\end{equation}
\section{The mass matrix\label{sec3}}

The conditions for the minimum of the potential are obtained from the
vanishing of the first derivatives at the minimum $\left.
  \frac{\partial V}{\partial \phi _{i}}\right| _{\min}=0$, with the
condition that the matrix of the second derivatives at the minimum:
$\left. \frac{\partial ^{2}V}{\partial \phi _{i}\partial \phi
    _{j}}\right|_{\min}$ is positive definite. Therefore
\begin{multline}
  \left. \frac{\partial V}{\partial \phi _{i}}\right| _{\left\langle
      0\right| \phi _{i}\left| 0\right\rangle } =\left\langle 0\right|
  \mu _{1}^{2}\frac{
    \partial }{\partial \phi _{i}}A+\mu _{2}^{2}\frac{\partial
  }{\partial \phi _{i}}B+2\lambda _{1}A\frac{\partial }{\partial \phi
    _{i}}A+2\lambda _{2}B \frac{\partial }{\partial \phi _{i}}B
  +\lambda _{3}B\frac{\partial A}{\partial \phi _{i}}+\lambda
  _{3}A\frac{
    \partial B}{\partial \phi _{i}}\\
  +\lambda _{4}D^{\prime }\frac{\partial C^{\prime }}{\partial \phi
    _{i}} +\lambda _{4}C^{\prime }\frac{\partial D^{\prime }}{\partial
    \phi _{i}} +\lambda _{5}\left[ C^{\prime }\frac{\partial
    }{\partial \phi _{i}} C^{\prime }+D^{\prime }\frac{\partial
    }{\partial \phi _{i}}D^{\prime }\right] \left| 0\right\rangle =0
\end{multline}
from which after some simplifications two non trivial equations are
obtained
\begin{equation}
  \mu _{1}^{2}+\lambda _{1}v_{1}^{2}+2\lambda_{T} \,v_{2}^{2}=0\;\text{ or }\;
  v_{1}=0,\quad \mu _{2}^{2}+\lambda _{2}v_{2}^{2}+2\lambda_{T}
  v_{1}^{2}=0\;\text{ or }\; v_{2}=0,  \label{v1}
\end{equation}
where
\begin{equation}
  \lambda_{T} \equiv \left( \lambda _{3}+\lambda _{4}+\lambda
    _{5}\right) .  \label{R1}
\end{equation}

The mass matrix elements are obtained from the equation
\begin{equation}
  M_{ij}^{2}=\frac{1}{2}\left. \frac{\partial ^{2}V}{\partial \phi
      _{i}\partial \phi _{j}}\right| _{\phi _{3}=\frac{v_{1}}{\sqrt{2}},\phi _{7}=\frac{v_{2}}{\sqrt{2}}},
        \label{nondiag}
\end{equation}
and the explicit form of the matrix of the second derivatives reads
\begin{multline}
  \frac{\partial ^{2}V}{\partial \phi _{j}\partial \phi _{i}} =\left(
    \mu _{1}^{2}+2\lambda _{1}A+\lambda _{3}B\right) \frac{\partial
    ^{2}A}{\partial \phi _{j}\partial \phi _{i}}+\left( \mu
    _{2}^{2}+2\lambda _{2}B+\lambda _{3}A\right) \frac{\partial
    ^{2}B}{\partial \phi _{j}\partial \phi _{i}} +2\lambda
  _{1}\frac{\partial A}{\partial \phi _{j}}\frac{\partial A}{
\partial \phi _{i}}+2\lambda _{2}\frac{\partial B}{\partial \phi _{j}}\frac{
\partial B}{\partial \phi _{i}}\\
+\lambda _{3}\left( \frac{\partial A}{
\partial \phi _{j}}\frac{\partial B}{\partial \phi _{i}}+\frac{\partial B}{
\partial \phi _{j}}\frac{\partial A}{\partial \phi _{i}}\right)
+\lambda _{5}\left( \frac{\partial C^{\prime }}{\partial \phi _{j}}\frac{
\partial C^{\prime }}{\partial \phi _{i}}+\frac{\partial D^{\prime }}{
\partial \phi _{j}}\frac{\partial D^{\prime }}{\partial \phi _{i}}\right)
+\lambda _{4}\left( \frac{\partial C^{\prime }}{\partial \phi _{j}}\frac{
\partial D^{\prime }}{\partial \phi _{i}}+\frac{\partial D^{\prime }}{
\partial \phi _{j}}\frac{\partial C^{\prime }}{\partial \phi
_{i}}\right)\\
+\left( \lambda _{4}D^{\prime }+\lambda _{5}C^{\prime }\right) \frac{
\partial ^{2}C^{\prime }}{\partial \phi _{j}\partial \phi _{i}}+\left(
\lambda _{4}C^{\prime }+\lambda _{5}D^{\prime }\right) \frac{\partial
^{2}D^{\prime }}{\partial \phi _{j}\partial \phi _{i}}.
\end{multline}
Using Eqs.~(\ref{v1}) the 16 non vanishing matrix elements are
\begin{equation}
  \begin{split}
M_{11}^{2} &=M_{22}^{2}=-\frac{1}{2}\left( \lambda _{4}+\lambda _{5}\right)
v_{2}^{2},\quad M_{33}^{2}=2\lambda
_{1}v_{1}^{2},\quad M_{44}^{2}=-\lambda _{5}v_{2}^{2},   \\
M_{55}^{2} &=M_{66}^{2}=-\frac{1}{2}\left( \lambda _{4}+\lambda _{5}\right)
v_{1}^{2},\quad M_{77}^{2}=2\lambda
_{2}v_{2}^{2},\quad M_{88}^{2}=-\lambda _{5}v_{1}^{2},  
\\
M_{15}^{2} &=M_{51}^{2}=M_{26}^{2}=M_{62}^{2}=\frac{1}{2}\left( \lambda
_{4}+\lambda _{5}\right) v_{1}v_{2},   \\
M_{37}^{2} &=M_{73}^{2}=\left( \lambda _{3}+\lambda _{4}+\lambda
_{5}\right) v_{1}v_{2},\quad M_{48}^{2}=M_{84}^{2}=\lambda _{5}v_{1}v_{2}.
\label{mxel}
\end{split}
\end{equation}

\subsection*{Diagonalization of the mass matrix}

The Higgs masses and the Higgs mass-eigenstates are obtained after a
suitable diagonalization of the matrix in Eq.~(\ref{nondiag}). The
diagonalization of the matrix whose elements are given in
Eq.~(\ref{mxel}) is performed in two steps.

A block ordering is performed and a diagonalization of each block is
carried out. The blocks are obtained by means of the application of
two consecutive unitary transformations $U_{1}=U_{1}^{\dagger }$ and
$U_{2}=U_{2}^{\dagger }$
\begin{equation}
  \left( M_{ij}^{2}\right) _{B}=U_{2}U_{1}M_{ij}^{2}U_{1}^{\dagger
  }U_{2}^{\dagger }.
\end{equation}
The non vanishing matrix elements of the unitary transformations are
\begin{gather*}
  \left( U_{1}\right) _{11}=\left( U_{1}\right) _{25}=\left( U_{1}\right)
  _{33}=\left( U_{1}\right) _{44}=\left( U_{1}\right) _{52}=\left(
    U_{1}\right) _{66}=\left( U_{1}\right) _{77}=\left( U_{1}\right) _{88}=1,\\
  \left( U_{2}\right) _{11}=\left( U_{2}\right) _{22}=\left( U_{2}\right)
  _{33}=\left( U_{2}\right) _{47}=\left( U_{2}\right) _{55}=\left(
    U_{2}\right) _{66}=\left( U_{2}\right) _{74}=\left( U_{2}\right) _{88}=1.
\end{gather*}
After carrying out both transformations, the $8\times 8$ matrix
becomes
\begin{equation}
\left( M_{ij}^{2}\right) _{B}=\left(
\begin{array}{llllllll}
M_{11}^{2} & M_{15}^{2} & 0 & 0 & 0 & 0 & 0 & 0 \\
M_{51}^{2} & M_{55}^{2} & 0 & 0 & 0 & 0 & 0 & 0 \\
0 & 0 & M_{33}^{2} & M_{37}^{2} & 0 & 0 & 0 & 0 \\
0 & 0 & M_{73}^{2} & M_{77}^{2} & 0 & 0 & 0 & 0 \\
0 & 0 & 0 & 0 & M_{22}^{2} & M_{26}^{2} & 0 & 0 \\
0 & 0 & 0 & 0 & M_{62}^{2} & M_{66}^{2} & 0 & 0 \\
0 & 0 & 0 & 0 & 0 & 0 & M_{44}^{2} & M_{48}^{2} \\
0 & 0 & 0 & 0 & 0 & 0 & M_{84}^{2} & M_{88}^{2}
\end{array}
\right) .  \label{blo}
\end{equation}
The matrix in Eq.~(\ref{blo}), is ready to easily perform the total
diagonalization
\begin{equation}
\left( M_{ij}^{2}\right) _{B}=\left(
\begin{array}{ll}
M_{11}^{2} & M_{15}^{2} \\
M_{51}^{2} & M_{55}^{2}
\end{array}
\right) \oplus \left(
\begin{array}{ll}
M_{33}^{2} & M_{37}^{2} \\
M_{73}^{2} & M_{77}^{2}
\end{array}
\right) \oplus \left(
\begin{array}{ll}
M_{22}^{2} & M_{26}^{2} \\
M_{62}^{2} & M_{66}^{2}
\end{array}
\right) \oplus \left(
\begin{array}{ll}
M_{44}^{2} & M_{48}^{2} \\
M_{84}^{2} & M_{88}^{2}
\end{array}
\right) .  \label{sub}
\end{equation}
The next step is to perform the diagonalization of each of the
submatrices in Eq.~(\ref{sub}).

\section{Higgs mass-eigenstates basis\label{sec4}}

Let us now proceed to relate the gauge states with the mass
eigenstates.

The scalar fields in Eq.~(\ref{geig}) can be represented
as
\begin{equation}
\Phi _{1}=\left(
\begin{array}{l}
\phi _{1}^{+} \\
\phi _{1}^{0}
\end{array}
\right) ;\,\,\,\,\Phi _{2}=\left(
\begin{array}{l}
\phi _{2}^{+} \\
\phi _{2}^{0}
\end{array}
\right)
\end{equation}
where
\begin{equation}
  \phi _{1}^{+}=\phi _{1}+i\phi _{2},\quad \phi _{2}^{+}=\phi _{5}+i\phi
  _{6},\quad \phi _{1}^{0}=\phi _{3}+i\phi _{4},\quad \phi _{2}^{0}=\phi
  _{7}+i\phi _{8},
\end{equation}
and
\begin{equation}
  \phi _{3}=\frac{v_{1}}{\sqrt{2}}+h_{1},\,\;\phi _{4}= \eta
    _{1},\quad \phi _{7}=\frac{v_{2}}{\sqrt{2}}+h_{2},\quad
  \phi_{8}=\eta _{2}.
\end{equation}
Now, the physical fields (mass eigenstates) and the Goldstones
(massless eigenstates) are obtained from the gauge eigenstates by a
unitary transformation that diagonalizes the corresponding submatrices
Eq.~(\ref{sub}), in the following way
\begin{equation}
\left(
\begin{array}{l}
H^{0} \\
h^{0}
\end{array}
\right) =U_{\alpha }\left(
\begin{array}{l}
h_{1} \\
h_{2}
\end{array}
\right) ,\quad \left(
\begin{array}{l}
G^{+} \\
H^{+}
\end{array}
\right) =U_{\beta }\left(
\begin{array}{l}
\phi _{1}^{+} \\
\phi _{2}^{+}
\end{array}
\right) ,\quad \left(
\begin{array}{l}
G^{0} \\
A^{0}
\end{array}
\right) =U_{\gamma }\left(
\begin{array}{l}
\eta _{1} \\
\eta _{2}
\end{array}
\right) ,
\end{equation}
where
\begin{equation}
U_{\alpha }=\left(
\begin{array}{cc}
\cos \alpha & \sin \alpha \\
-\sin \alpha & \cos \alpha
\end{array}
\right) ,\quad U_{\alpha }^{\dagger }U_{\alpha }=I
\end{equation}
and $U_{\beta } = U_{\gamma }$ have the same form as $U_{\alpha }$.
$\alpha $ is the mixing angle between the neutral states $\phi
_{1}^{0}$ and $\phi _{2}^{0}$, i.e.,  $\phi_{3}$ and $\phi_{7}$ , the
$\beta $ angle is the one between the charged states, and $\gamma$ is
related with the CP-Odd states $\phi _{4}$ and $ \phi _{8}$.
\begin{equation}
  \tan \alpha =\frac{y}{1+\sqrt{1+y^{2}}} ,\quad y=\tan 2\alpha =\frac{%
    \left( \lambda _{3}+\lambda _{4}+\lambda _{5}\right) v_{1}v_{2}}{\left(
      \lambda _{1}v_{1}^{2}-\lambda _{2}v_{2}^{2}\right)} ,\quad
  -\frac{\pi }{2}<\alpha <\frac{%
    \pi }{2},  \label{tan}
\end{equation}
and
\begin{equation}
  \tan \beta =\frac{v_{2}}{v_{1}},\quad v^{2}=\left(
    v_{1}^{2}+v_{2}^{2}\right),  \quad 0<\beta <\frac{\pi }{2} ,\quad
  \gamma =\beta. 
    \end{equation}   
The resulting physical particles in the Higgs-sector are: two
CP-even-neutral Higgs scalars ($H^{0},h^{0}$), one CP-odd neutral
Higgs scalar ($A^{0}$), two charged Higgs bosons ($H^{\pm }$), and
three Goldstone-bosons ($G^{\pm },G^{0}$) that contribute to the mass
generation of the gauge vector bosons $W^{\pm }$ and $Z^{0}$,
respectively.
\subsection*{The mass formulas}
After the complete diagonalization, we obtain the following relations:
\begin{enumerate}
\item The mass eigenvalues for ($H^{0},h^{0}$) are
\begin{equation}
  M_{H^{0},h^{0}}^{2}=\lambda _{1}v_{1}^{2}+\lambda _{2}v_{2}^{2}\pm \sqrt{\left(
\lambda _{1}v_{1}^{2}-\lambda _{2}v_{2}^{2}\right) ^{2}+\left(
v_{1}v_{2}\lambda_{T} \right) ^{2}}>0,  \label{EquivK}
\end{equation}

\item The eigenvalues for the mass eigenstates $H^{\pm }$ and $G^{\pm
  }$ are
\begin{equation}
  M_{G^{\pm }}^{2}=0\;,\quad M_{H^{\pm }}^{2}=-\frac{1}{2}\left(
    \lambda _{4}+\lambda _{5}\right) v^{2}>0,  \label{Hmm}
\end{equation}
\item
Finally, the mass eigenvalues for $G^{0}$ and $A^{0}$ are
\begin{equation}
  M_{G^{0}}^{2}=0,\quad M_{A^{0}}^{2}=-\lambda _{5}v^{2}>0,
\label{A0}
\end{equation}
\end{enumerate}
As expected, after the electroweak symmetry breaking (EWSB) the eight
components of the two complex isodoublet fields are transformed into:
two charged Higgs bosons $H^{\pm }$, three neutral Higgs bosons
$H^{0},h^{0},A^{0}$, and three massless Goldstone fields $G^{0},G^{\pm
}$ (which are transformed into the longitudinal components of the
gauge bosons $W^{\pm }$ and $Z^{0}$).  At this level, the values of $
M_{A^{0}}$ and $M_{H^{\pm }}$ are not related to the parameters
$\lambda _{1}$, $\lambda _{2}$ and $\lambda _{3}$.  This means that,
apparently, there is a complete independence between
the $A^{0}$, $H^{\pm }$  and  the $h^{0}$, $ H^{0}$, which is not all true.\\
As in the Standard Model, the values of the quartic couplings are not
fixed by the model.  To proceed as in the SM~\cite{KJ}, to determine
the Higgs masses, one has to consider two important physical
principles.  The vacuum stability constrains the values for the
quartic couplings.  To have a complete view, we invert the former
equations to express the quartic parameters in terms of the masses of
the Higgs fields.
\begin{gather}
  \lambda _{1}=\frac{1}{2v_{1}^{2}}\left( M_{H^{0}}^{2}\cos ^{2}\alpha
    +M_{h^{0}}^{2}\sin ^{2}\alpha \right), \quad\lambda
  _{2}=\frac{1}{2v_{2}^{2}}%
  \left( M_{H^{0}}^{2}\sin ^{2}\alpha +M_{h^{0}}^{2}\cos ^{2}\alpha
  \right) \label{32}\\
\lambda _{3}=\frac{\left( M_{H^{0}}^{2}-M_{h^{0}}^{2}\right) }{2v_{1}v_{2}}%
\sin 2\alpha +2\frac{M_{H^{\pm }}^{2}}{v^{2}},\quad 
\lambda _{4}=\frac{M_{A^{0}}^{2}-2M_{H^{\pm }}^{2}}{v^{2}}\quad 
\lambda _{5}=-\frac{M_{A^{0}}^{2}}{v^{2}},\label{32b}\\
  \lambda_{T}=\left(\lambda _{3}+\lambda _{4}+\lambda
    _{5}\right)=\frac{\left(
      M_{H^{0}}^{2}-M_{h^{0}}^{2}\right) }{2v_{1}v_{2}}\sin 2\alpha. \label{34}
\end{gather}

\subsection*{Particular cases}
 
 To obtain Eq.~({\ref{32}})-({\ref{34}}) we have considered the
 following relations: Since Eq.~({\ref{EquivK}}) is equivalent to
\begin{equation}
  M_{H^{0},h^{0}}^{2}=\left[ \lambda _{1}v_{1}^{2}+\lambda _{2}v_{2}^{2}\pm
    \left( \lambda _{1}v_{1}^{2}-\lambda _{2}v_{2}^{2}\right) \sqrt{1+y^{2}}%
  \right] 
\end{equation}
and
\begin{equation}
\sqrt{1+y^{2}}=\sqrt{1+\left( \tan 2\alpha \right) ^{2}}=\frac{1}{\cos
2\alpha },
\end{equation}
 we obtain from Eq.~({\ref{tan}})
\begin{equation}
  \sin 2\alpha =\frac{\left( \lambda _{3}+\lambda _{4}+\lambda _{5}\right)
    v_{1}v_{2}}{\sqrt{\left( \lambda _{1}v_{1}^{2}-\lambda _{2}v_{2}^{2}\right)
      ^{2}+\left( \lambda _{3}+\lambda _{4}+\lambda _{5}\right) ^{2}\left(
        v_{1}v_{2}\right) ^{2}}} \label{s1}
\end{equation}
as well as
\begin{equation}
  \cos 2\alpha =\frac{\lambda _{1}v_{1}^{2}-\lambda _{2}v_{2}^{2}}{\sqrt{%
      \left( \lambda _{1}v_{1}^{2}-\lambda _{2}v_{2}^{2}\right) ^{2}+\left(
        \lambda _{3}+\lambda _{4}+\lambda _{5}\right) ^{2}\left( v_{1}v_{2}\right)
      ^{2}}} \label{c1}
\end{equation} 
and
\begin{equation}
\frac{\left( M_{H^{0}}^{2}-M_{h^{0}}^{2}\right) \cos 2\alpha }{2}=\left(
\lambda _{1}v_{1}^{2}-\lambda _{2}v_{2}^{2}\right)
,\quad\,\frac{M_{H^{0}}^{2}+M_{h^{0}}^{2}}{2}%
=\left[ \lambda _{1}v_{1}^{2}+\lambda _{2}v_{2}^{2}\right]. 
\end{equation}
With these equations it is easy to obtain Eq.~(\ref{32}).
\vspace{0.25in}\\
a.- In the case when the mixing angle is $\alpha=0$, i.e., $\lambda_{T}
=0$, $\lambda _{3}>0$,
\begin{gather}
  \lambda _{1}=\frac{1}{2v_{1}^{2}}M_{H^{0}}^{2},\quad\lambda _{2}=
  \frac{1}{2v_{2}^{2}}M_{h^{0}}^{2},\quad\lambda _{3}=2\left(
    \frac{M_{H^{\pm }}}{v}\right) ^{2},  \\
  \lambda _{4}=\left( \frac{M_{A^{0}}}{v}\right) ^{2}-2\left( \frac{M_{H^{\pm
        }}}{v}\right) ^{2},\quad\lambda _{5}=-\left( \frac{M_{A^{0}}}{v}
  \right) ^{2}, 
\end{gather}
and
\begin{equation}
M_{H^{0}}^{2}-M_{h^{0}}^{2}=2\left( \lambda _{1}v_{1}^{2}-\lambda
_{2}v_{2}^{2}\right) .
\end{equation}

The parameters $\mu _{1},\mu _{2}$ in the potential Eq.~(\ref{ec1}) are
related to the neutral Higgs particles in a very simple way, similar
to the one between the parameters $\lambda $ and $\mu $ in the SM.
\begin{equation}
  -2\mu _{1}^{2}=M_{H^{0}}^{2}=2\lambda
  _{1}v_{1}^{2},\quad -2\mu
  _{2}^{2}=M_{h^{0}}^{2}=2\lambda _{2}v_{2}^{2}.
\end{equation}
and the vevs satisfy
\begin{equation}
  v^{2}=\left( v_{1}^{2}+v_{2}^{2}\right) =-\left( \frac{\mu _{2}^{2}}{\lambda
      _{2}}+\frac{\mu _{1}^{2}}{\lambda _{1}}\right) =\frac{1}{2\lambda
    _{1}\lambda _{2}}\left( \lambda _{1}M_{h^{0}}^{2}+\lambda
    _{2}M_{H^{0}}^{2}\right) .
\end{equation}
In this particular case, each Higgs particle is associated with a
specific parameter $\lambda _{1}$, $\lambda _{2}$, $\lambda _{3}$,
$\lambda _{5}$.
\begin{equation}
  M_{H^{0}}=v_{1}\sqrt{2\lambda
    _{1}},\quad M_{h^{0}}=v_{2}\sqrt{2\lambda _{2}},\quad M_{H^{\pm
    }}=\frac{v}{\sqrt{2}}\sqrt{\lambda _{3}},\quad
  M_{A^{0}}=v\sqrt{\left| \lambda _{5}\right| }.
\end{equation}
The degeneracy in the masses
$M_{H^{0}}$, $M_{h^{0}}$ implies that $\lambda _{1}v_{1}^{2}-\lambda
_{2}v_{2}^{2}=0$.\\[7pt]
b.- In the case when the mixing angle is $\alpha=\pi/2$, $\lambda_{T} =0$
\begin{gather}
  \lambda _{1}=\frac{1}{2v_{1}^{2}}M_{h^{0}}^{2},\quad\lambda _{2}=
  \frac{1}{2v_{2}^{2}}M_{H^{0}}^{2},\quad\lambda _{3}=2\left(
    \frac{M_{H^{\pm }}}{v}\right) ^{2},  \\
  \lambda _{4}=\left( \frac{M_{A^{0}}}{v}\right) ^{2}-2\left( \frac{M_{H^{\pm
        }}}{v}\right) ^{2},\quad\lambda _{5}=-\left( \frac{M_{A^{0}}}{v}
  \right) ^{2}.
\end{gather}
The parameters $\mu _{1},\mu _{2}$ in Eq.~(\ref{v1}) become:
\begin{equation}
  -2\mu _{1}^{2}=M_{h^{0}}^{2}=2\lambda
  _{1}v_{1}^{2},\quad -2\mu
  _{2}^{2}=M_{H^{0}}^{2}=2\lambda _{2}v_{2}^{2}.
\end{equation}
and 
\begin{equation}
  v^{2} =\frac{1}{2\lambda
    _{1}\lambda _{2}}\left( \lambda _{1}M_{H^{0}}^{2}+\lambda
    _{2}M_{h^{0}}^{2}\right) .
\end{equation}\
As in the former case, each Higgs particle is associated with a
specific parameter  $\lambda _{i}$, and ($H^{0}, h^{0}$) interchange places.
\begin{equation}
  M_{h^{0}}=v_{1}\sqrt{2\lambda
    _{1}},\quad M_{H^{0}}=v_{2}\sqrt{2\lambda _{2}},\quad M_{H^{\pm
    }}=\frac{v}{\sqrt{2}}\sqrt{\lambda _{3}},\quad
  M_{A^{0}}=v\sqrt{\left| \lambda _{5}\right| }.
\end{equation}
In this section, we have considered the main features of various special cases for the
parameter $\alpha $ where a decoupling of the Higgs bosons take place, 
and the case where the masses of the CP-even neutral particles coincide.

\section{ Vacuum stability constrains\label{sec5}}
\subsection*{Bounds due to the positive mass-values}
Due to the fact that the masses are positive, from the previous
results, one gets information for the allowed values of the $\lambda
_{i}$ parameters.
\begin{equation}
  \lambda _{1}>0,\quad \lambda _{2}>0,\quad \left( \lambda
    _{4}+\lambda _{5}\right) <0,\ \ \lambda _{5}<0, \ \ \lambda _{4}<\left\vert
    \lambda _{5}\right\vert .\label{m1}
\end{equation}
and Eq.~(\ref{EquivK}) implies that
\begin{equation}
  \lambda _{1}\lambda _{2}>\frac{1}{4}\left( \lambda _{3}+\lambda _{4}+\lambda
    _{5}\right) ^{2}.\label{m2}
\end{equation}
In terms of the masses, the conditions in Eq.~(\ref{m1}) and
Eq.~(\ref{m2}) become trivial
\begin{gather}
  \lambda _{1}=\frac{1}{2v_{1}^{2}}\left[ M_{H^{0}}^{2}\cos ^{2}\alpha
    +M_{h^{0}}^{2}\sin ^{2}\alpha \right] >0,  \label{l1}\\
  \lambda _{2}=\frac{1}{2v_{2}^{2}}\left[ M_{H^{0}}^{2}\sin ^{2}\alpha
    +M_{h^{0}}^{2}\cos ^{2}\alpha\right] >0,  \label{l2}\\
\lambda _{1}\lambda _{2}-\frac{1}{4}\left( \lambda _{3}+\lambda _{4}+\lambda
_{5}\right) ^{2}=\frac{1}{4v_{1}^{2}v_{2}^{2}}M_{H^{0}}^{2}M_{h^{0}}^{2}>0.
\end{gather}
To improve previous information about the allowed values of the
quartic couplings and therefore for the masses, we have explored the
consecuences of considering the vacuum stability conditions (VSC),
through the method of the Lagrange multipliers.

\subsection*{Lagrangian multipliers method and the VSC}
\textit{Considering one restriction}: Let us introduce the variables
$x_{i}$ and the parameters $b_i$, defined as~\footnote{To make the
  discussion more transparent we introduced a different
  parametrization of the Higgs potential in this section.}
\begin{gather}
  x_{1}=\left| \Phi _{1}\right| ^{2},\quad x_{2}=\left| \Phi _{2}\right|
  ^{2},\,\,x_{3}=\frac{1}{2}\left( \Phi _{1}^{\dagger }\Phi _{2}+\Phi
    _{2}^{\dagger }\Phi _{1}\right) ,\,\,\,\,x_{4}=\frac{1}{2i}\left( \Phi
    _{1}^{\dagger }\Phi _{2}-\Phi _{2}^{\dagger }\Phi _{1}\right) ,\label{27}\\
  b_{11}=\lambda _{1},\quad b_{22}=\lambda _{2},\quad b_{33}=\left(
    \lambda _{4}+\lambda _{5}\right) ,\quad b_{44}=\left( \lambda _{4}-\lambda
    _{5}\right) ,\quad b_{12}=\lambda _{3},
\end{gather}
the potential in Eq.~(\ref{ec1}) becomes $V=V_{0}+x_{1}^{2}F_{0},$
where $V_{0}=\mu _{1}^{2}x_{1}+\mu _{2}^{2}x_{2}$, and
\begin{equation}
F_{0}=b_{11}+b_{12}\xi _{2}+b_{22}\xi _{2}^{2}+b_{33}\xi _{3}^{2}+b_{44}\xi
_{4}^{2},\quad\xi_{i}=\frac{x_{i}}{x_{1}},\;i=2,3,4. 
\end{equation}
Using the Cauchy-Scwartz inequality
\begin{equation}
  \left| \Phi _{1}^{\dagger }\Phi _{2}\right| ^{2}\leq \Phi _{1}^{\dagger
  }\Phi _{1}\Phi _{2}^{\dagger }\Phi _{2}=\left| \Phi _{1}\right| ^{2}\left|
    \Phi _{2}\right| ^{2}, \label{cauchy} 
\end{equation}
we obtain the condition
\begin{equation}
  f\left( \xi _{2},\xi _{3},\xi _{4}\right) =\xi _{3}^{2}+\xi _{4}^{2}-\xi
  _{2}\leq 0.\label{rcon}
\end{equation}
We now introduce the Lagrange multiplier $\Lambda _{1}$ in the quartic
sector of the potential~\footnote{The vacuum stability constrains for
  the Higgs potential do not depend on the quadratic part of the Higgs
  potential and for this reason we do not include it in our
  discussion.} related to the condition imposed by the Cauchy-Schwarz
inequality Eq.~(\ref{rcon})
\begin{equation}
  F\left( \xi _{2},\xi _{3},\xi _{4},\Lambda _{1}\right)=F_{0}\left( \xi
    _{2},\xi _{3},\xi _{4}\right) +\Lambda _{1}\,f\left( \xi _{2},\xi _{3},\xi
    _{4}\right)= 
  b_{11}+\left( b_{12}-\Lambda _{1}\right) \xi _{2}+b_{22}\xi
  _{2}^{2}+\left( b_{33}+\Lambda _{1}\right) \xi _{3}^{2}+\left(
    b_{44}+\Lambda _{1}\right) \xi _{4}^{2}
\label{RA1}
\end{equation}
and apply the stability (positivity) condition in Eq.~(\ref{RA1}) after
obtaining the derivatives
\begin{equation}
  \frac{\partial F}{\partial \xi _{2}}=\frac{\partial F}{\partial \xi _{3}}=%
  \frac{\partial F}{\partial \xi _{4}}=\frac{\partial F}{\partial \Lambda _{1}}%
  =0, 
\end{equation}
to evaluate the minimum value
for $F\left( \xi _{2},\xi _{3},\xi _{4},\Lambda _{1}\right)$ in the
region of interest.

The following equations are to be solved.
 \begin{equation}
   2b_{22}\xi _{2}+b_{12}-\Lambda _{1}=0,\quad 2\left( b_{33}+\Lambda
     _{1}\right) \xi _{3}=0,\quad 2\left( b_{44}+\Lambda _{1}\right)
   \xi _{4}=0,\quad \xi _{3}^{2}+\xi _{4}^{2}-\xi _{2}=0.  \label{EqA2}	
\end{equation}
There are two solutions denoted by $A_{i}$, $i=1,2$, where $\xi
_{2}=\xi_{3}^{2}+\xi _{4}^{2}$, and
\begin{equation}
  \Lambda _{1}=2b_{22}\xi _{2}+b_{12},\quad 
  \left( b_{33}+2b_{22}\xi _{2}+b_{12}\right) \xi _{3}=0,\quad  \left(
    b_{44}+2b_{22}\xi _{2}+b_{12}\right) \xi _{4}=0, \label{EqA3}
\end{equation}
where the stability condition becomes
\begin{equation}
  \left. F\left( \xi _{2},\xi _{3},\xi _{4},\Lambda _{1}\right) \right| _{\min }=\left. (b_{11}-b_{22}\xi
    _{2}^{2}+\left( b_{33}+\Lambda _{1}\right) \xi _{3}^{2}+\left(
      b_{44}+\Lambda _{1}\right) \xi _{4}^{2}\right)| _{\min }>0.\label{cA}
\end{equation}
There is another solution, case $B$, in which $\Lambda _{1}=0$, \quad
and \quad $\xi _{3}^{2}+\xi _{4}^{2}-\xi_{2}\leq 0$. Where
\begin{equation}
  \left. F\left( \xi _{2},\xi _{3},\xi _{4}\right)\right| _{\min
  }=\left. (b_{11}+ b_{12} \xi _{2}+b_{22}\xi
    _{2}^{2}+b_{33} \xi _{3}^{2}+b_{44} \xi _{4}^{2}\right)| _{\min }>0. \label{cB}
\end{equation}
In the first solution $A_{1}$ we consider:
\begin{equation}
  \xi _{4}=0,\quad \xi _{3}\neq 0,\quad \xi _{2}=-\frac{1}{%
    2b_{22}}\left( b_{12}+b_{33}\right)=\xi_{3}^{2}.
\end{equation}
The conditions and implications for a minimum in the region of interest are:
\begin{equation}
b_{12}+b_{33}\leq 0,\quad b_{22}>0,\quad \lambda
_{2}>0,\quad \lambda _{3}+\lambda _{4}+\lambda _{5}\leq 0.	
\end{equation}
In the second solution $A_{2}$
\begin{equation}
\quad \xi _{4}\neq 0,\quad \left( b_{33}+2b_{22}\xi _{2}+b_{12}\right) \xi
_{3}=0,\quad b_{44}+2b_{22}\xi_{2}+b_{12}=0.
\end{equation}
We have two possibilities, $\xi _{3}=0$,\quad and \quad $\xi _{3}\neq 0$, and 
the existence of a minimum requires, if $\xi _{3}=0$  
\begin{equation}
  \xi _{2}=\xi _{4}^{2},\quad \xi _{2}=-\frac{\left(
      b_{12}+b_{44}\right) }{2b_{22}},\quad 
  b_{22}>0,\quad b_{12}+b_{44}<0,\quad \lambda_{2}>0,\quad \lambda _{3}+\lambda
  _{4}-\lambda _{5}\leq 0.	
\end{equation}
If $\xi _{3}\neq 0$, the implications are
\begin{equation}
\xi_{2}=-\frac{\left( b_{12}+b_{44}\right) }{2b_{22}}=
-\frac{\left( b_{12}+b_{33}\right) }{2b_{22}},\quad \left(
  b_{33}-b_{44}\right) \xi _{3}=0, 
\quad \lambda_{5}=0,\quad \lambda _{3}+\lambda _{4}\leq 0. 	
\end{equation}
In case $B$, the equations to solve are \quad $2b_{22}\xi
_{2}+b_{12}=0$,\quad $2b_{33}\xi_{3}=0$, \quad $2b_{44}\xi _{4}=0$,
and the conditions to have a minimum with its implications are
\begin{equation}
  \xi _{2}=-\frac{b_{12}}{2b_{22}}>0,\quad  b_{12}\leq0,\quad
  b_{22}>0,\quad \lambda _{3}<0. 
\end{equation}
Now we apply the stability condition in Eq.~(\ref{cA}) and obtain in
cases $A_{1}$ and $A_{2}$ 
\begin{align}
4b_{22}b_{11}\geq \left( b_{12}+b_{33}\right)
^{2}&\Rightarrow -2\sqrt{\lambda_{1}\lambda _{2}}<\left( \lambda
  _{3}+\lambda _{4}+\lambda _{5}\right),\\
  4b_{11}b_{22} >\left( b_{12}+b_{44}\right) ^{2}&\Rightarrow
  -2\sqrt{\lambda _{1}\lambda _{2}} <\left( \lambda _{3}+\lambda
    _{4}-\lambda _{5}\right),\\
  b_{33}=b_{44},\quad \lambda _{5}=0&\Rightarrow
  -2\sqrt{\lambda _{1}\lambda _{2}} <\left( \lambda _{3}+\lambda
    _{4}\right).
\end{align}
Now, in case $B$ and Eq.~(\ref{cB}), the result is
\begin{equation}
  4b_{22}b_{11}>\left( b_{12}\right) ^{2}\Rightarrow \lambda _{3}>-2\sqrt{%
    \lambda _{1}\lambda _{2}}.
\end{equation}
Performing the second derivative
\begin{equation}
  F\left( \xi _{2},\xi _{3},\xi _{4},\Lambda _{1}\right) =b_{11}+\left(
    b_{12}-\Lambda _{1}\right) \xi _{2}+b_{22}\xi _{2}^{2}+\left( b_{33}+\Lambda
    _{1}\right) \xi _{3}^{2}+\left( b_{44}+\Lambda _{1}\right) \xi _{4}^{2}.	
\end{equation}
We obtain 
\begin{equation}
  \frac{\partial}{\partial\xi _{2}}\left( \frac{\partial F}{\partial\xi
      _{2}}\right) =\frac{\partial}{\partial\xi _{2}}
  \left( 2b_{22}\xi _{2}+b_{12}-\Lambda _{1}\right) =2b_{22}>0.	
\end{equation}
After imposing the stability condition, with one restriction we obtain 
these new boundary values for the quartic couplings:
\begin{equation}
  -2\sqrt{\lambda _{1}\lambda _{2}}<\left( \lambda _{3}+\lambda _{4}+\lambda
    _{5}\right), \quad
  -2\sqrt{\lambda _{2}\lambda _{1}}\leq \lambda _{3}.
\end{equation}
\textit{Considering now two restrictions}: Following the same method
as in the former case, considering now the conditions
\begin{equation}
  x_{3}^{2}+x_{4}^{2}\leq x_{1}x_{2},\quad x_{1}+x_{2}-v^{2}=0	
\end{equation}
and two Lagrange multipliers, the function in consideration is:
\begin{equation}
  F_{2c}\left( x_{1},x_{2},x_{3},x_{4},\Lambda _{1},\Lambda _{2}\right)=
  b_{11}x_{1}^{2}+b_{22}x_{2}^{2}+b_{33}x_{3}^{2}+b_{44}x_{4}^{2}+b_{12}x_{1}x_{2}
  +\Lambda _{1}\left( x_{3}^{2}+x_{4}^{2}-x_{1}x_{2}\right) +\Lambda
  _{2}\left( x_{1}+x_{2}-v^{2}\right)   \label{Req2-1}
\end{equation}
The new equations to solve are
\begin{align}
2b_{11}x_{1}+b_{12}x_{2}+\Lambda _{2}-\Lambda _{1}x_{2}
&=0,\quad 2b_{22}x_{2}+b_{12}x_{1}+
\Lambda _{2}-\Lambda _{1}x_{1}=0, \nonumber  \\
\left( b_{33}+\Lambda _{1}\right) x_{3}
&=0,\quad \left( \,b_{44}+\Lambda
_{1}\right) x_{4}=0.  \label{Req2-2} \\
\Lambda _{1}\left( x_{3}^{2}+x_{4}^{2}-x_{1}x_{2}\right) 
&=0,\quad \Lambda _{2}\left(
x_{1}+x_{2}-v^{2}\right) =0.  \nonumber
\end{align}
For  $\left. F_{2c}\left( x_{1},x_{2},x_{3},x_{4},\Lambda _{1},\Lambda
_{2}\right) \right| _{\min }$%
\begin{equation}
x_{1}=\left( \frac{b_{12}-\Lambda _{1}-2b_{22}}{4b_{22}b_{11}-(b_{12}-%
\Lambda _{1})^{2}}\right) \Lambda
_{2}>0,\quad x_{2}=\left( \frac{%
b_{12}-\Lambda _{1}-2b_{11}}{4b_{22}b_{11}-(b_{12}-\Lambda _{1})^{2}}\right)
\Lambda _{2}>0.  \label{Req2-3}
\end{equation}
and 
\begin{equation}
\left. F_{2c}\left( x_{1},x_{2},x_{3},x_{4},\Lambda _{1},\Lambda _{2}\right)
\right| _{\min }=\left. b_{11}x_{1}^{2}+b_{22}x_{2}^{2}+\left(
b_{12}-\Lambda _{1}\right) x_{1}x_{2}\right| _{\min }  \label{Req2-4}
\end{equation}
The solutions satisfy the following requirements: 
\begin{equation}
\frac{\left. F_{2c}\left( x_{1},x_{2},x_{3},x_{4},\Lambda _{1},\Lambda
_{2}\right) \right| _{\min }}{\Lambda _{2}^{2}}=\frac{\left(b_{11}+b_{22} -b_{12}+\Lambda
_{1}\right) }{ 4b_{22}b_{11}-(b_{12}-\Lambda
_{1})^{2}}>0,  \label{Req2-5}
\end{equation}
with
\begin{equation}
\Lambda _{2}=\frac{4b_{22}b_{11}-(b_{12}-\Lambda _{1})^{2}}{2\left(
b_{12}-\Lambda _{1}-b_{22}-b_{11}\right) }v^{2}<0.  \label{Req2-6}
\end{equation}
The requirements in Eqs.~(\ref{Req2-3}) imply
\begin{equation}
  \begin{array}{l}
 \left( b_{12}-\Lambda _{1}\right) ^{2}-4b_{11}b_{22}
<0,\quad b_{12}-\Lambda _{1}-2b_{22}<0, \\
 \left( b_{12}-\Lambda
_{1}-b_{11}-b_{22}\right) <0,\quad b_{12}-\Lambda _{1}-2b_{11}<0.
\end{array}
\label{Req2-7}
\end{equation}

The aditional solutions that come from Eqs.~(\ref{Req2-2}) yield
\[
\Lambda _{1}=-b_{33},\quad \Lambda_{1}=-b_{44},\quad b_{33}=b_{44}.
\]
Then
\[
\left( \lambda _{3}+\lambda _{4}\pm \left| \lambda _{5}\right| \right)
^{2}<4\lambda _{1}\lambda _{2},\quad\lambda _{3}+\lambda _{4}\pm \left|
\lambda _{5}\right| <\lambda _{1}+\lambda _{2},\quad\lambda
_{3}+\lambda _{4}\pm \left| \lambda _{5}\right| <2\lambda
_{2},\quad \lambda _{3}+\lambda _{4}\pm \left| \lambda
_{5}\right| <2\lambda _{1}
\]
and
\[
-2\sqrt{\lambda _{1}\lambda _{2}}<\left( \lambda _{3}+\lambda _{4}\pm \left|
\lambda _{5}\right| \right) ,\quad \lambda _{3}+\lambda _{4}\pm
\left| \lambda _{5}\right| <2\lambda
_{1},\quad \lambda _{3}+\lambda _{4}\pm \left|
\lambda _{5}\right| <2\lambda _{2}.
\]
Thus we obtain previous results plus
\begin{equation}
\lambda _{1}+\lambda _{2}>\lambda _{3}+\lambda _{4}\pm \left|
\lambda _{5}\right| ,\quad \lambda _{1}+\lambda _{2}>\lambda _{3}+\lambda
_{4}.
\end{equation}

\subsection*{Bounds from extreme stability conditions }
Let us now determine the behavior of the quartic couplings in the case
where the quartic Higgs potential $V_4$ has its lowest possible
value.  In correspondence with Eq.~(\ref{ec1}) using the notation of
Eq.~(\ref{27}) when $x_{1}=v_{1}^{2}$, $x_{2}=v_{2}^{2}$ the $V_{4}$
can be simplified as follows
\[
V_{4}=\lambda _{1}x_{1}^{2}+\lambda _{2}x_{2}^{2}+\lambda
_{T}x_{1}x_{2}=\left( \sqrt{\lambda _{1}}x_{1}-\sqrt{\lambda _{2}}x_{2}\right)
^{2}+\left( \lambda_{T}+2\sqrt{\lambda _{1}\lambda _{2}}\right)
x_{1}x_{2}\geq 0.
\]
In the Extreme case, the condition to be satisfied is $V_{4}=0$, then
\[
\sqrt{\frac{\lambda _{1}}{\lambda _{2}}}=\frac{x_{2}}{x_{1}}=\frac{v_{2}^{2}}{v_{1}^{2}},
\quad \lambda _{T}=-2\sqrt{\lambda _{1}\lambda
_{2}},\quad \Rightarrow \quad \frac{\lambda _{1}}{\lambda _{2}}=\frac{v_{2}^{4}}{v_{1}^{4}}=\left( \tan
\beta \right) ^{4}.
\]
In this case the Higgs masses become
\[
M_{H^{0}}=\left\{ 
\begin{array}{ll}
\left( 4\lambda _{1}\lambda _{2}\right) ^{1/4}v ,&\quad \lambda _{1}\neq \lambda
_{2},\quad v_{1}\neq v_{2} ,\\ 
\left(2\lambda\right) ^{1/2}v ,&\quad \lambda _{1}=\lambda _{2}=\lambda ,\quad v_{1}=v_{2}=v/%
\sqrt{2}.
\end{array}
\right. 
\]
In another interesting case, which is the Semi-extreme case, the
$V_{4}=\left( \sqrt{\lambda _{1}}x_{1}-\sqrt{\lambda _{1}}%
  x_{1}\right) ^{2}>0$, and
\[
\frac{\lambda _{1}}{\lambda _{2}}\neq \left( \tan \beta \right)
^{4},\quad \lambda_{T}=-2\sqrt{\lambda _{1}\lambda _{2}},
\]
the $M_{H^{0}}$ becomes
\[
M_{H^{0}}=
\begin{array}{ll}
\sqrt{2}\left( \lambda _{1}v^{2}+\left( \lambda _{2}-\lambda _{1}\right)
v_{2}^{2}\right) ^{1/2} & \
\end{array}. \
\]
In both cases
\[
M_{h^{0}}=0,\quad M_{H^{\pm }}=\left( \frac{1}{2}%
\left| \lambda _{4}+\lambda _{5}\right| \right)
^{1/2}v,\quad M_{A^{0}}=\left| \lambda
_{5}\right| ^{1/2}v.
\]
\subsection*{Numerical evaluation of the Higgs masses in terms of
  $\tan \beta$ }

\begin{figure}[ht]
\centering
\includegraphics[width=0.45\linewidth]{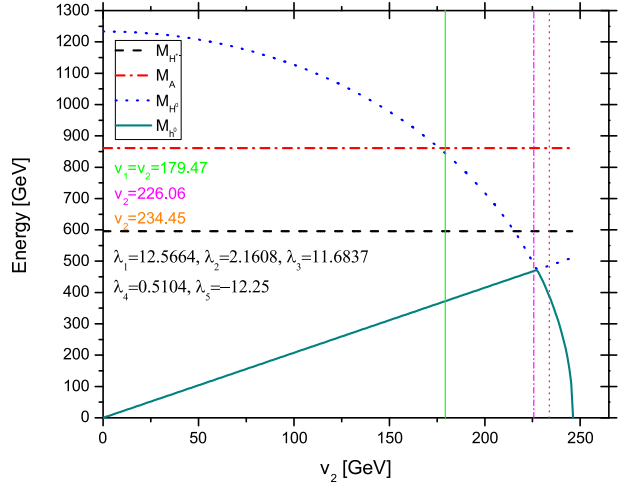}
\caption{The $v_2$ dependence of the Higgs masses
for various sets of quartic couplings.\label{fig:1}}
\end{figure}

In general, according to Eqs.~\eqref{A0}-\eqref{32b} the mass
dependence on
$v_{1},v_{2},\,\,$or$\,\,\,v=\sqrt{v_{1}^{2}+v_{2}^{2}},$ can be
reformulated in terms of $v$ and $v_{2}.$ Therefore, with known $v$
one can plot those masses in terms of $v_{2\text{ }}$, and determine
their dependence on $v_{2}$ or $\tan \beta$ ($\tan \beta
=v_{2}/\sqrt{v^{2}-v_{2}^{2}}$), as in Fig.~\ref{fig:1}.

As the favoured Standard Model Higgs mass window is still open, and as
the 2HDM Higgses are nonstandard, almost none of the masses range has
yet been ruled out, one can explore several hypothetical situations
and analize the consequences of some published values for
$M_{H^{\pm}}$.

We will now proceed to numerically evaluate the Higgs masses under
different conditions for the quartic parameters at the energy scale
$E=M_{t}$, where $M_{t}$ is the mass of the quark top. First we will
reproduce the $M_{H^{\pm }}$ value given in~\cite{Stal}, considering
several sets of $\lambda $'s.  We choose $\lambda _{4}=-5.52,$ and
$\lambda _{5}=-6.0$ from many combinatios that give $M_{H^{\pm
  }}=609.1, M_{A^{0}}=621.7$. As we shall see, in the extreme case,
with one of the symmetries considered in~\cite{Maniatis} in which
$\lambda _{1}=\lambda _{2}=\lambda$, and $\tan \beta =1$, or $\tan
\beta =5$ all the Higgs masses acquire constant values and
$M_{h^{0}}=0$. With the previous $\lambda _{4}$, $\lambda _{5}$, we
obtain for $H^{0}$ values which are not ruled out experimentaly (for
the SM Higgs) according to~\cite{D0}, in the following way: ($\lambda
=0.485$, $\lambda _{3}=10.55$, $\lambda _{T}=-0.97$, $M_{H^{0}}=250$),
($\lambda =0.653$, $\lambda _{3}=10.21$, $\lambda _{T}=-1.30$,
$M_{H^{0}}=290$) and ($\lambda =1.94$, $\lambda _{3}=7.64$, $\lambda
_{T}=-3.88$, $M_{H^{0}}=600$).\

Though the masses do not depend on $\tan \beta$ at this energy scale, we
will explore, in the next section, their behavior and dependence on it
at higher energies scales.

Let us now classify the several cases, we will consider , in terms of
the different stability condiions for the $\lambda _{i}$ s

A.- Extreme case in which both equalities are satisfied 
\begin{equation}
\lambda _{T}=-2\sqrt{\lambda _{1}\lambda _{2}},\,\,\,\,\,\,\,\cup
\,\,\,\,\,\,\,\,\lambda _{1}/\lambda _{2}=\left( \tan \beta \right) ^{4} 
\end{equation}

B1.- Semi-extreme case: 
\begin{equation}
\lambda _{T}=-2\sqrt{\lambda _{1}\lambda _{2}},\,\,\,\,\,\,\lambda
_{1}/\lambda _{2}\neq \left( \tan \beta \right) ^{4} 
\end{equation}

B2.- Semi-extreme case: 
\begin{equation}
\lambda _{T}\neq -2\sqrt{\lambda _{1}\lambda _{2}},\,\,\,\,\,\,\lambda
_{1}/\lambda _{2}=\left( \tan \beta \right) ^{4} 
\end{equation}

C.- Lagrange inequality condition 
\begin{equation}
\lambda _{T}\geq -2\sqrt{\lambda _{1}\lambda _{2}} 
\end{equation}

D.- Yukawa - Universality condition 
\begin{equation}
\tan \beta =M_{t}/M_{b},\,\,\,\,\,\,\,\,\,\,\,\,\,\,\,\,g_{t}=g_{b} 
\end{equation}
The former cases will be combined with two additional conditions for
the quartic couplings:

Case 1 : $\lambda _{1}=\lambda _{2},$ and Case 2: $\lambda _{1}\neq \lambda
_{2}.$\\
In case 1A , 2A and 1B, all the masses have a constant value in the inerval $%
0\leq v_{2}\leq v$ due to their only dependence on $v,$ but in spite of the
an explicit independence of the masses on $v_{2}$ , $\tan \beta $ plays an
important role on the energy scale dependence of the masses, as we shall see
in the numerical solutions of the Renormalization Group Equations.which is
fixed. In case 2B, M$_{H^{0}}$ depends explicitly on $v_{2}$ and therefore
on $\tan \beta .$

To compare with the masses  $M_{H^{\pm }}=609,$ $M_{A}=621.7$ given in
Ref.\cite{Stal}, we consider three kinds of compositions of the $\lambda
_{i}\,\,\,i=1,...,5$  parameters as in a), b) and c), which produce seven
cases. The properties of these cases are to be analized in the following
section.\\[4pt]
a.-When $\left(\begin{array}{ccccc}
\lambda _{1} & \lambda _{2} & \lambda _{3} & \lambda _{4} & \lambda
_{5} \\ 
0.485 & 0.485 & 10.55 & -5.52 & -6.0
\end{array}\right)$ ,
which means that $\lambda _{T}=-2\sqrt{\lambda _{1}\lambda
  _{2}}=-0.97$ and $\left( \lambda _{1}/\lambda _{2}\right)
^{1/4}=1.0,$ we obtain
\[
\begin{array}{c}
\text{Case} \\ 
1A \\ 
1D
\end{array}
\begin{array}{ccc}
\tan \beta  & M_{h^{0}} & M_{H^{0}} \\ 
1.0 & 0 & 250 \\ 
41.2 & 0 & 250
\end{array}
\begin{array}{c}
v_{2} \\ 
179.47 \\ 
253.7
\end{array}
\]
b.- When $\left(\begin{array}{ccccc}
\lambda _{1} & \lambda _{2} & \lambda _{3} & \lambda _{4} & \lambda
_{5} \\ 
6.0 & 1/6 & 9.51 & -5.51 & -6.0%
\end{array}\right)$, i.e., $\lambda _{T}=-2\sqrt{\lambda _{1}\lambda _{2}}=-2.0$ and $\left(
\lambda _{1}/\lambda _{2}\right) ^{1/4}=2.45,$ 
we obtain
\[
\begin{array}{c}
\text{Case} \\ 
2A
\end{array}
\begin{array}{ccc}
\tan \beta  & M_{h^{0}} & M_{H^{0}} \\ 
2.45 & 0 & 358.9
\end{array}
\begin{array}{c}
v_{2} \\ 
234.9
\end{array}
\]
c.- When $\left(\begin{array}{ccccc}
\lambda _{1} & \lambda _{2} & \lambda _{3} & \lambda _{4} & \lambda
_{5} \\ 
12.0 & 3.0 & 11.5 & -5.5 & -6.0%
\end{array}\right)$ with $\lambda _{T}=0.0,-2\sqrt{\lambda _{1}\lambda _{2}}=-12.0$ and $\left(
\lambda _{1}/\lambda _{2}\right) ^{1/4}=1.4,$ we explore different $\tan
\beta $ cases 
\[
\begin{array}{c}
\text{Case} \\ 
2B2 \\ 
2C \\ 
2C \\ 
2D
\end{array}
\begin{array}{ccc}
\tan \beta  & M_{h^{0}} & M_{H^{0}} \\ 
1.4 & 507.6 & 717.9 \\ 
2.0 & 556.0 & 556.0 \\ 
5.0 & 243.9 & 609.6 \\ 
41.2 & 30.2 & 621.5
\end{array}
\begin{array}{c}
v_{2} \\ 
207.2 \\ 
227.0 \\ 
248.9 \\ 
253.7
\end{array}
\]
d.- Considering now smaller values for $\{M_{H^{\pm
}},M_{A}\}=\{253.8,240.8\}$, with interesting properties for the energy
range of validity of the 2HDM, with $\lambda _{T}=-2\sqrt{\lambda _{1}\lambda _{2}}=-0.97$ and $\left(
\lambda _{1}/\lambda _{2}\right) ^{1/4}=1.0$, for $\left(\begin{array}{ccccc}
\lambda _{1} & \lambda _{2} & \lambda _{3} & \lambda _{4} & \lambda
_{5} \\ 
0.48 & 0.48 & 1.03 & -1.1 & -0.9%
\end{array}\right)$
\[
\begin{array}{c}
\text{Case} \\ 
1A \\ 
1D
\end{array}
\begin{array}{ccc}
\tan \beta  & M_{h^{0}} & M_{H^{0}} \\ 
1.0 & 0 & 250 \\ 
41.2 & 0 & 250
\end{array}
\begin{array}{c}
v_{2} \\ 
179.5 \\ 
253.7
\end{array}
\]
e.- Finally, for even lower masses $\{M_{H^{\pm
  }},M_{A}\}=\{170.3,160.5\},$ which arise from
$\left(\begin{array}{ccccc} \lambda _{1} & \lambda _{2} & \lambda
  _{3} & \lambda _{4} & \lambda
  _{5} \\
  0.125 & 0.125 & 0.65 & -0.5 & -0.4%
\end{array}\right)$, where $\lambda _{T}=-2\sqrt{\lambda _{1}\lambda
_{2}}=-0.25$ and $\left( 
\lambda _{1}/\lambda _{2}\right) ^{1/4}=1.0,$ we obtain
\[
\begin{array}{c}
\text{Case} \\ 
1D,1B1
\end{array}
\begin{array}{ccc}
\tan \beta  & M_{h^{0}} & M_{H^{0}} \\ 
41.2 & 0 & 127
\end{array}
\begin{array}{c}
v_{2} \\ 
253.7
\end{array}
\]

\section{Triviality constrains. \label{sec6}}
\subsection*{Renormalization group equations}
In this section we explore the asymptotic behavior of the parameters
in the model, and their relations, through the Renormalization Group
Equations (RGE)~\cite{Inoue}.  The RGE are a powerful tool to
determine by the triviality principle, the energy bounds of the
parameters and the validity of the model.  In order to proceed in this
way, to numerically evaluate the energy dependence of the $\lambda
_{i}$ quartic couplings, it is necessary to consider the RGE of all
the parameters, i.e., the gauge group couplings $g_1$, $g_2$, $g_3$ of
the symmetry groups $U(1), SU(2), SU(3)$, the vacuum expectation
values $v_1$, $v_2$, and the Yukawa couplings of the top and the down
quark sectors $g_t$ and $g_d$ respectively Refs.~\cite{Arason}.

\begin{figure}[ht]
\centering
\includegraphics[width=.35\linewidth]{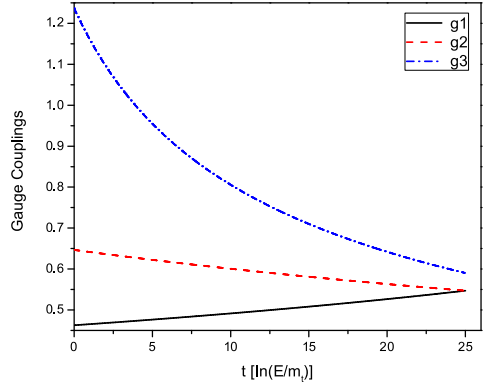}
\caption{The energy dependence of the gauge couplings in the
  2HDM.\label{fig:2}}
\end{figure}

\begin{figure}[ht]
\centering
\includegraphics[width=.4\linewidth]{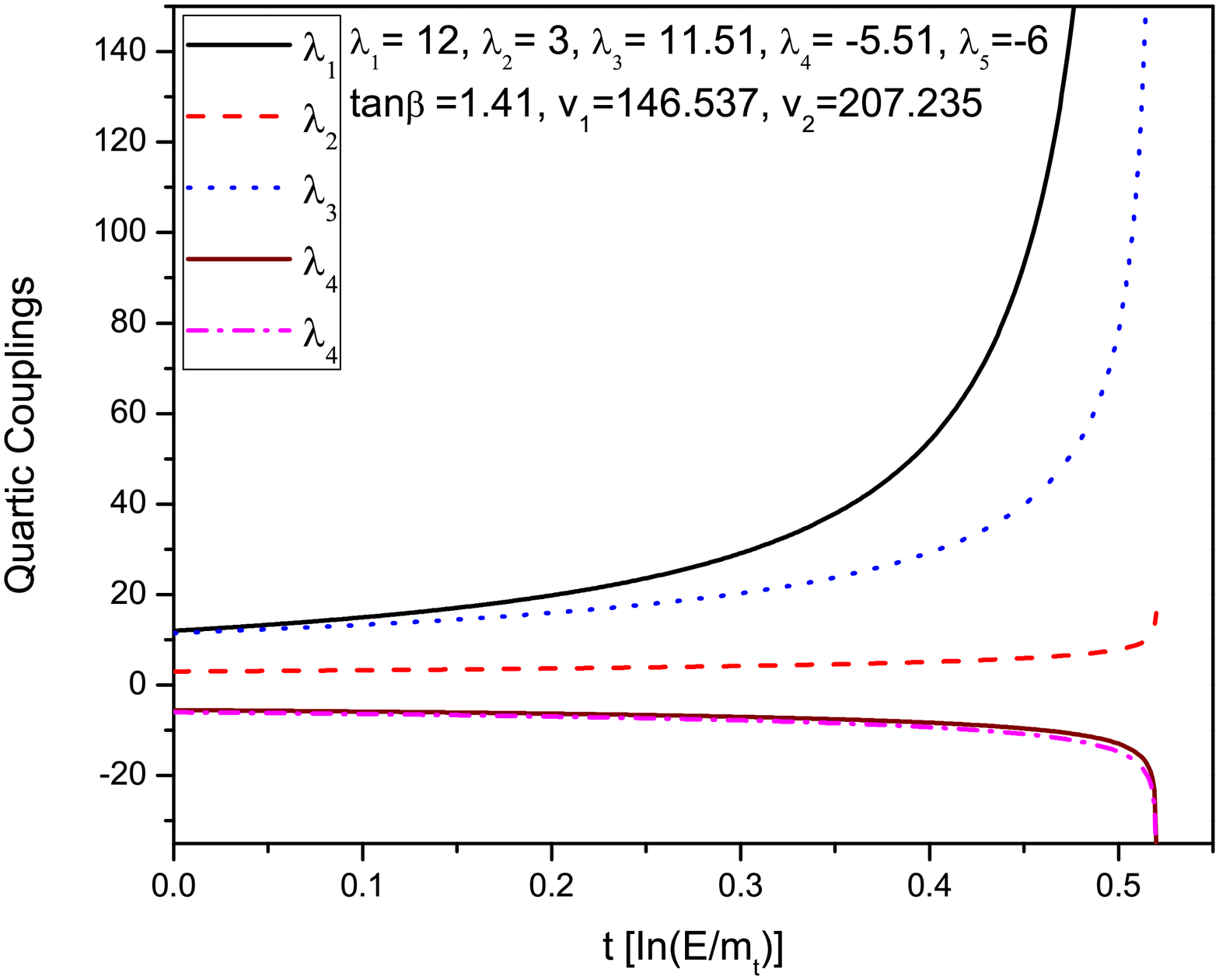}\hspace*{20pt}
\includegraphics[width=.4\linewidth]{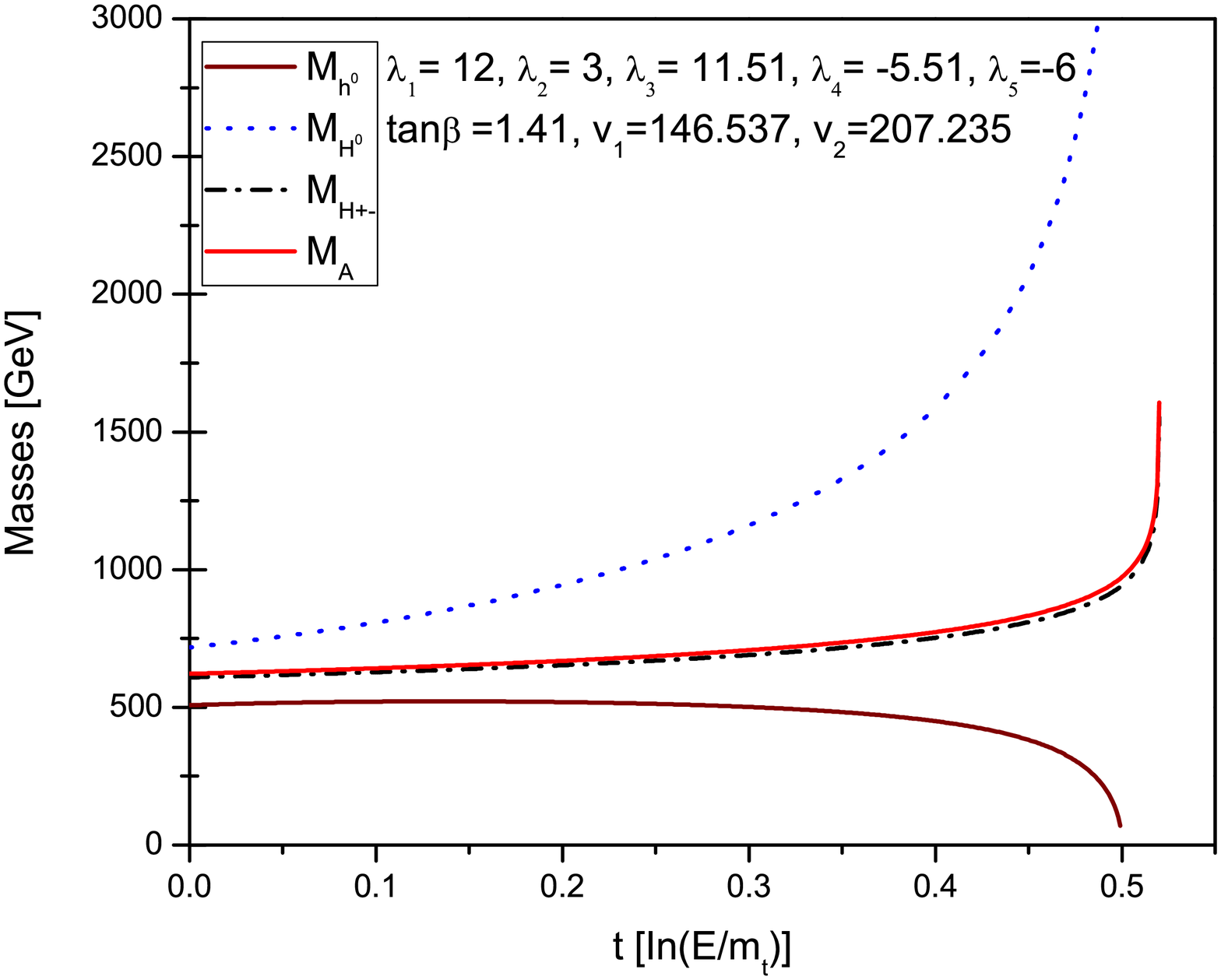}
\caption{The energy dependence of the quartic couplings and Higgs masses,
case 2B2, with $ \tan \beta =1.41$.\label{fig:3}}
\end{figure}

\begin{figure}[ht]
\centering
\includegraphics[width=.4\linewidth]{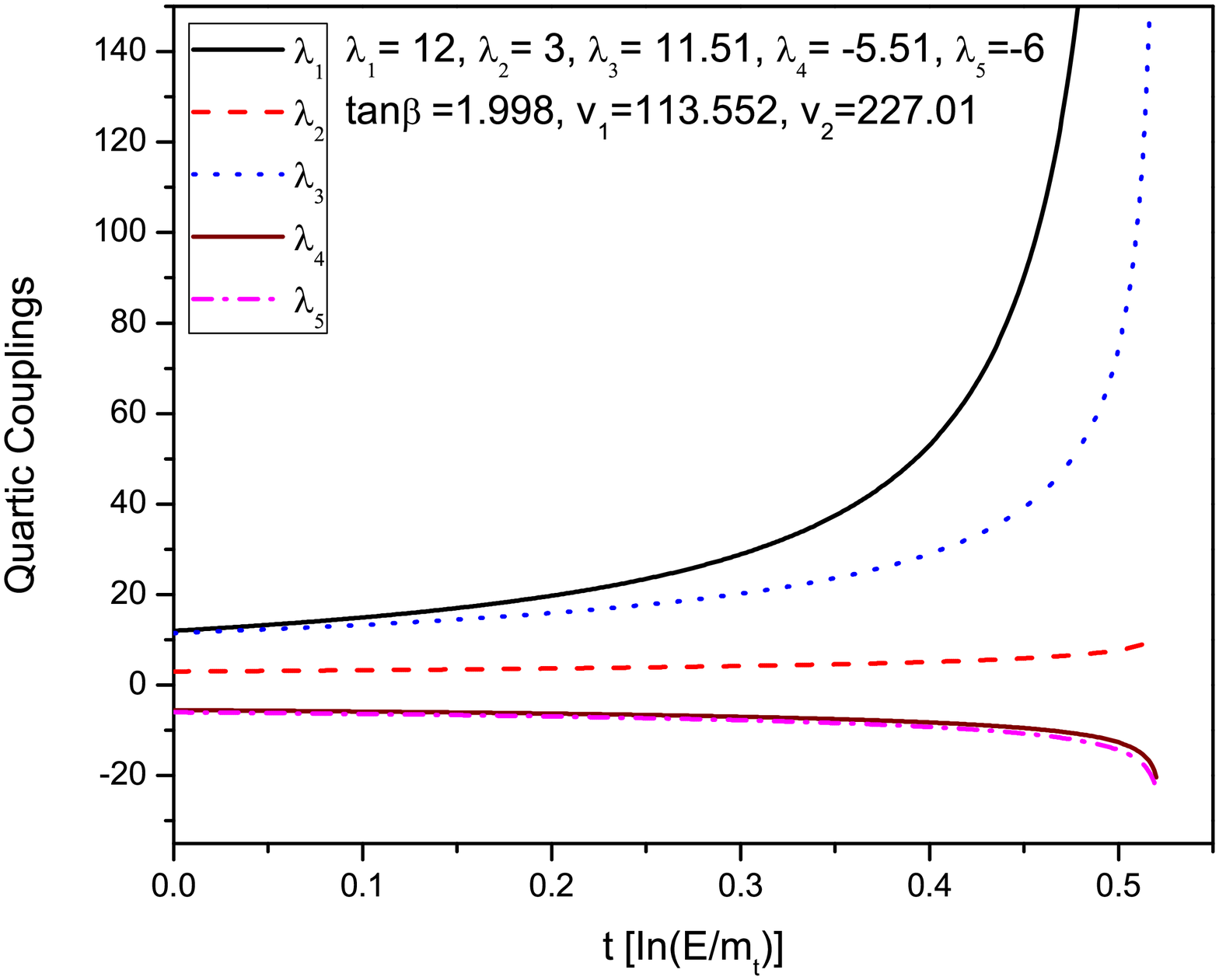}\hspace*{20pt}
\includegraphics[width=.4\linewidth]{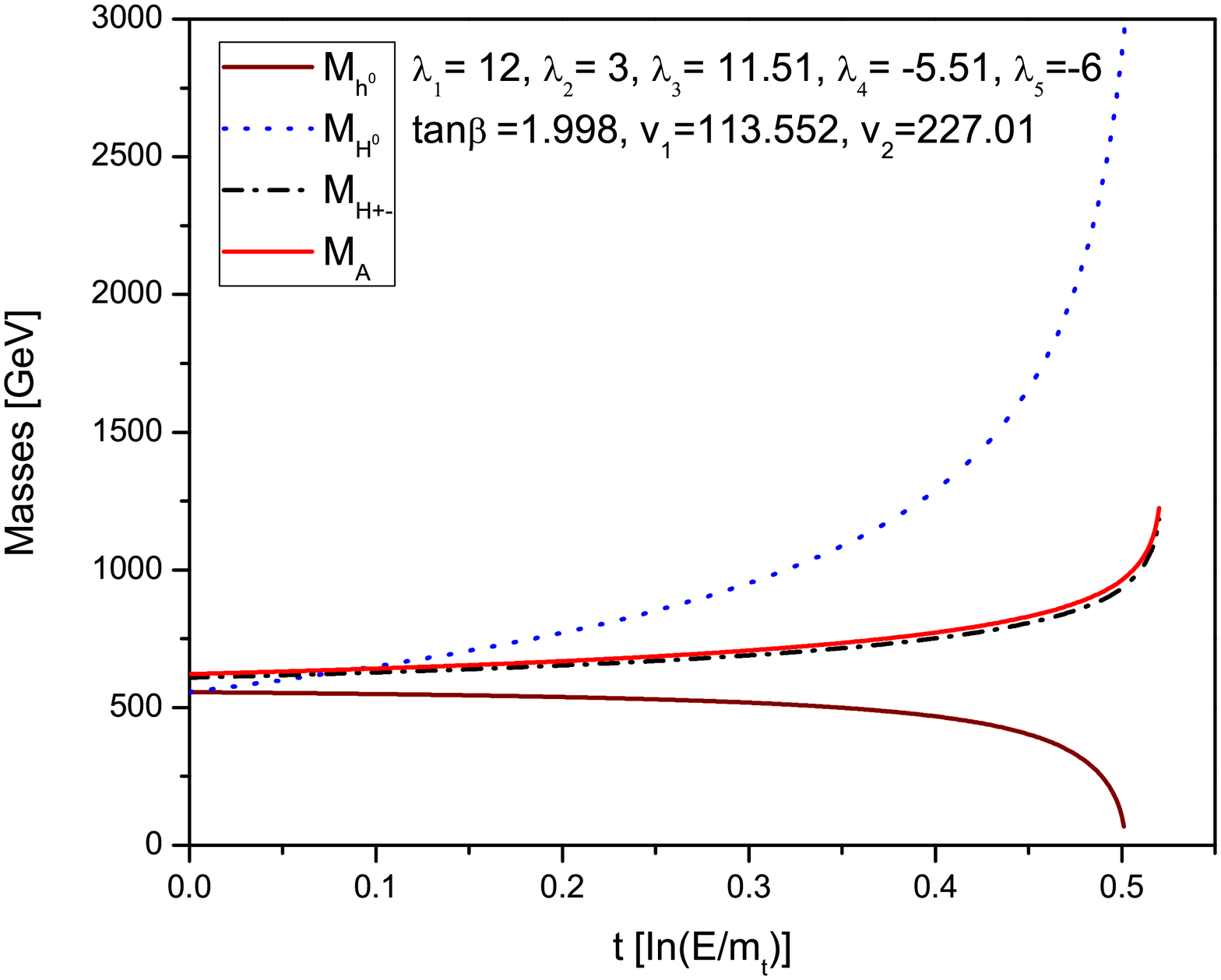}
\caption{The energy dependence of the quartic couplings and the Higgs masses,
case 2C with $\tan \beta =2.0$.\label{fig:4}}
\end{figure}

\begin{figure}[ht]
\centering
\includegraphics[width=.4\linewidth]{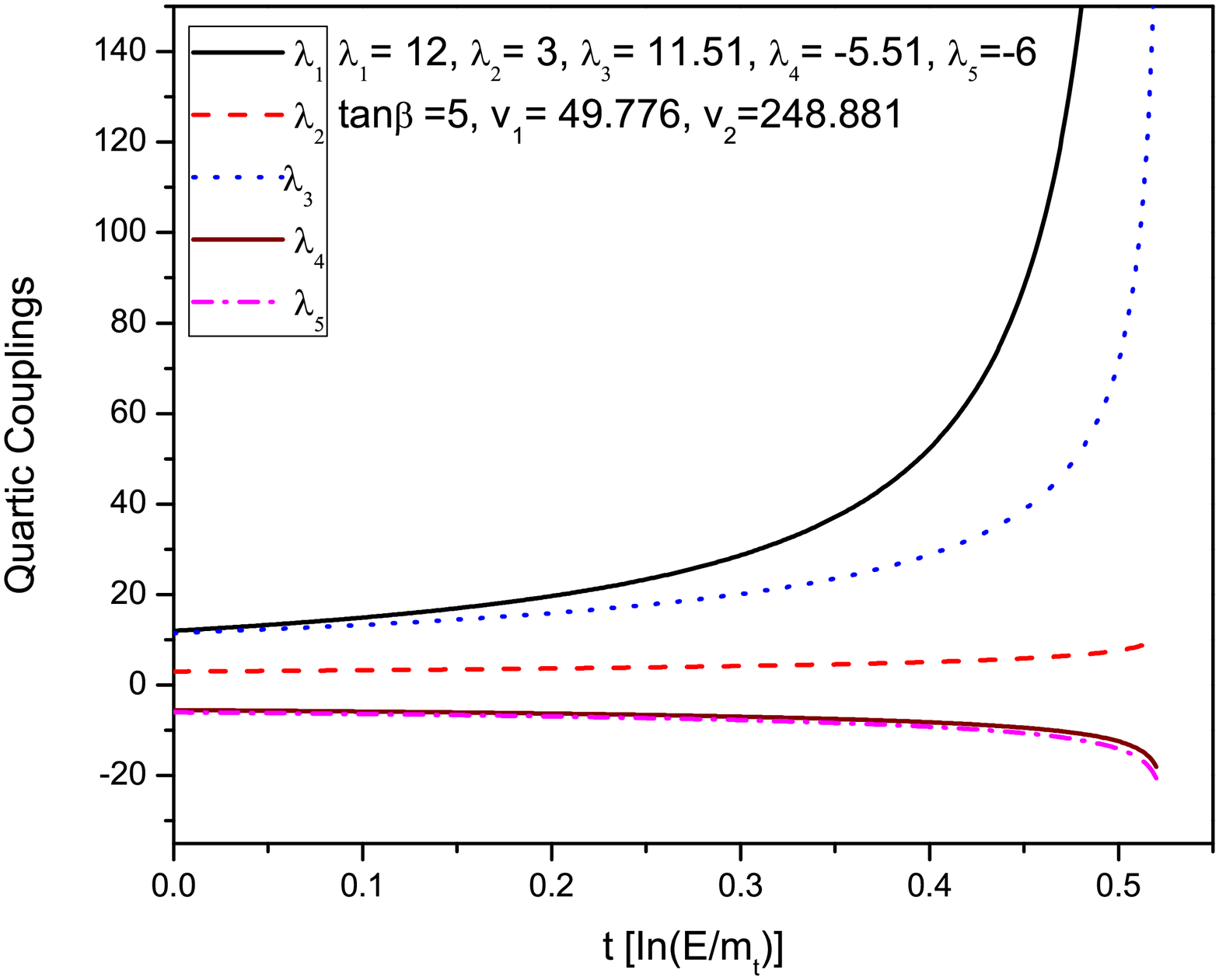}\hspace*{20pt}
\includegraphics[width=.4\linewidth]{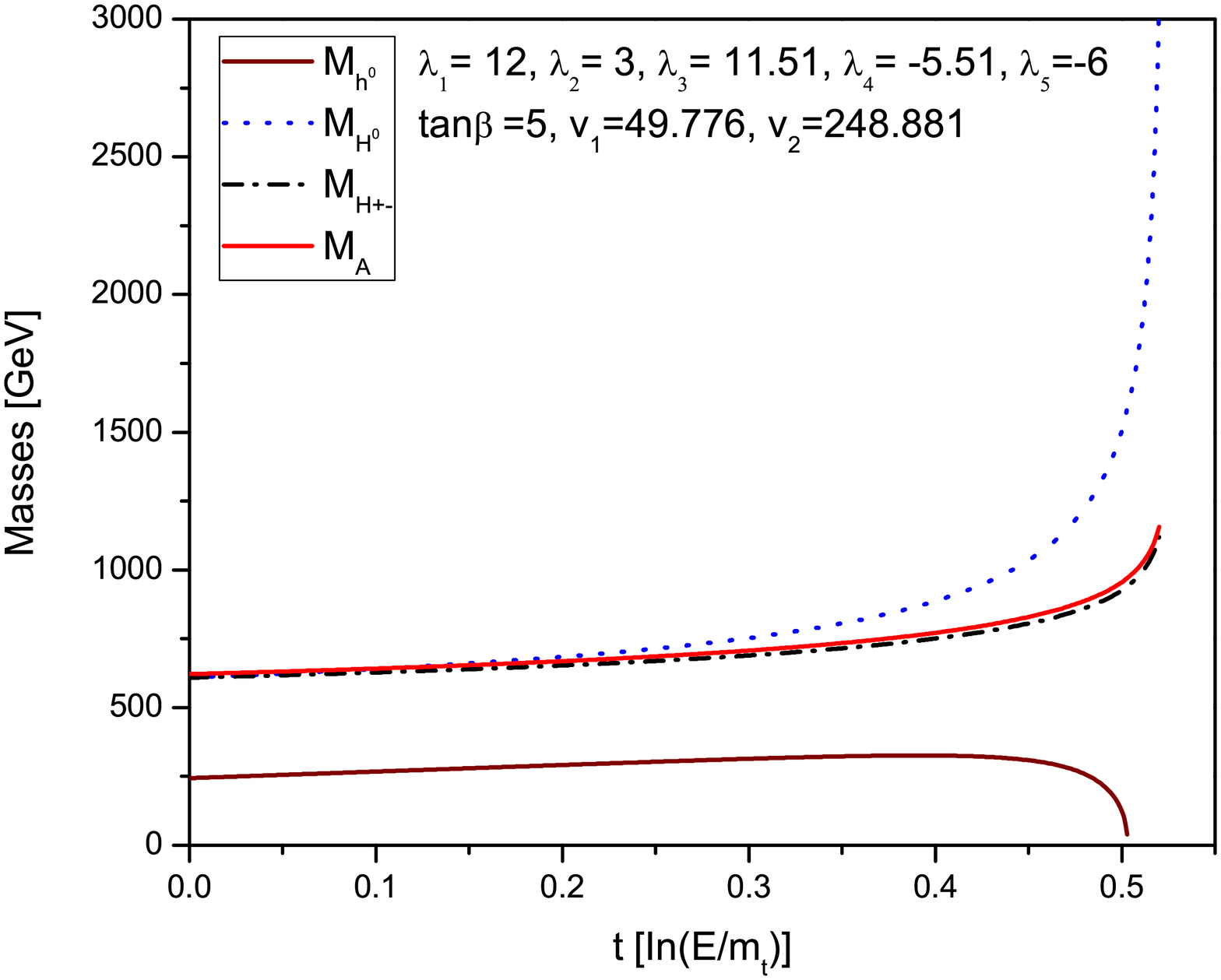}
\caption{The energy dependence of the quartic couplings and the Higgs masses,
case 2C with $\tan \beta =5.0$.\label{fig:5}}
\end{figure}

\begin{figure}[ht]
\centering
\includegraphics[width=.4\linewidth]{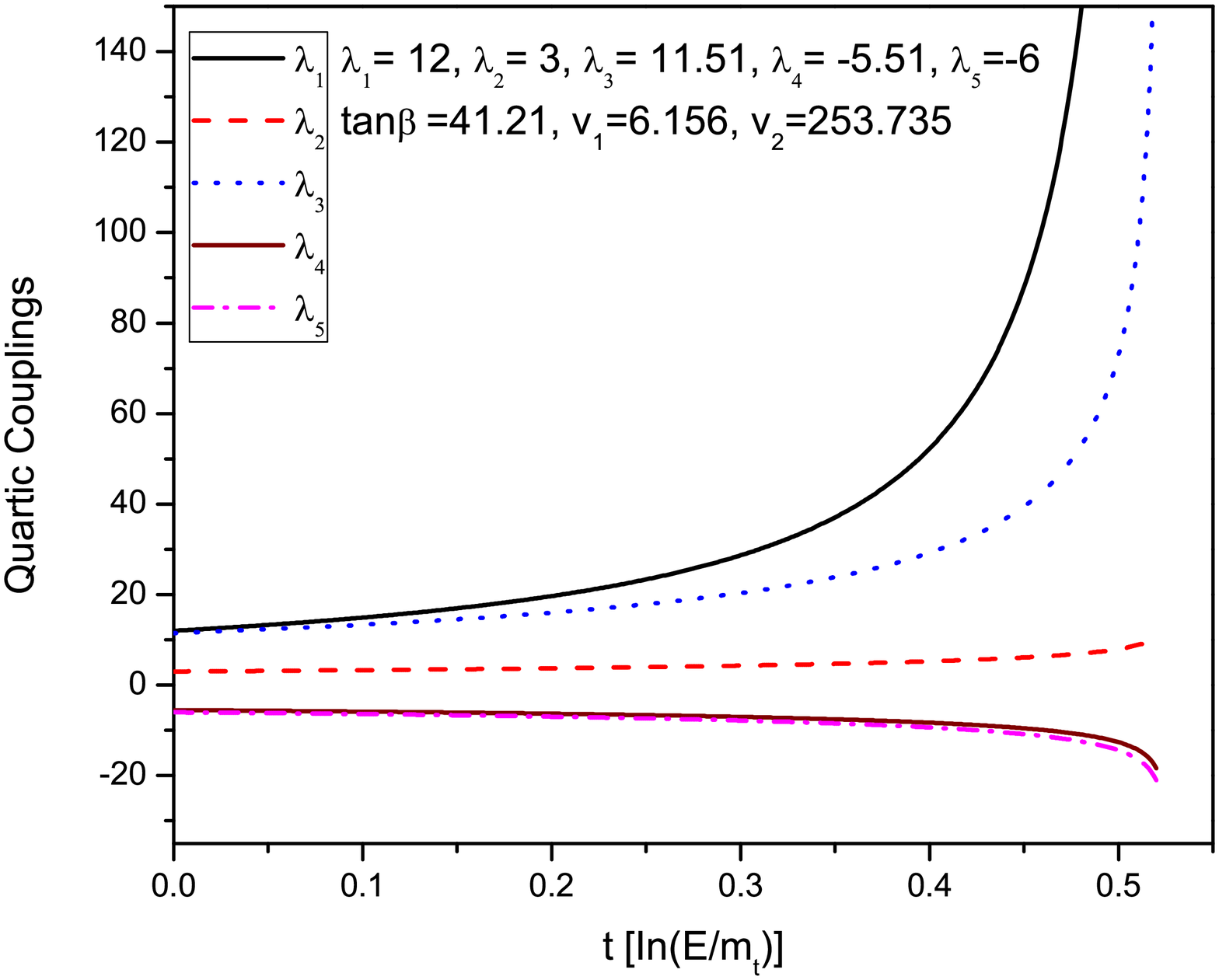}\hspace*{20pt}
\includegraphics[width=.4\linewidth]{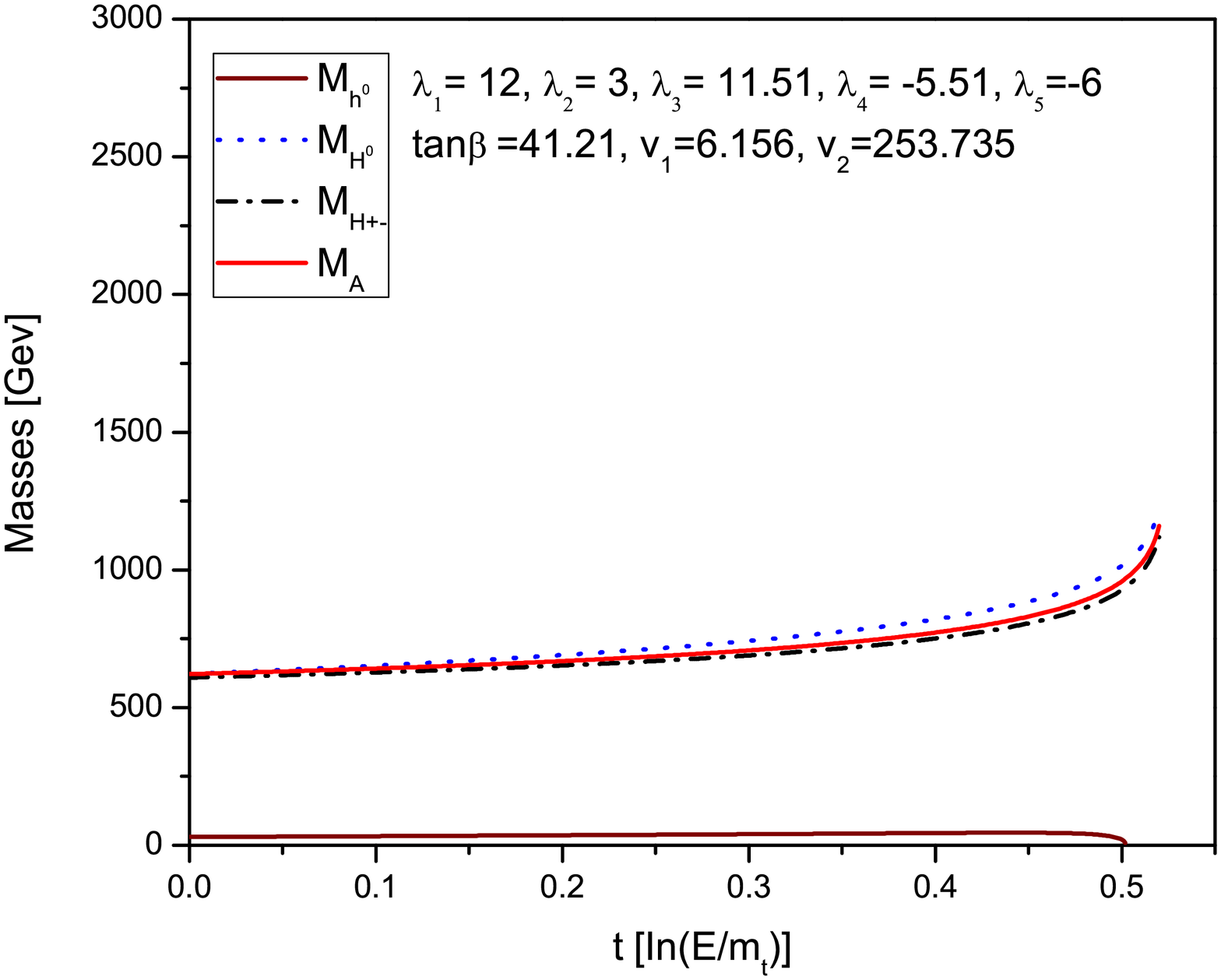}
\caption{The energy dependence of the quartic couplings and the Higgs masses,
case 2D with $\tan \beta =41.2$.\label{fig:6}}
\end{figure}

\begin{figure}[ht]
\centering
\includegraphics[width=.4\linewidth]{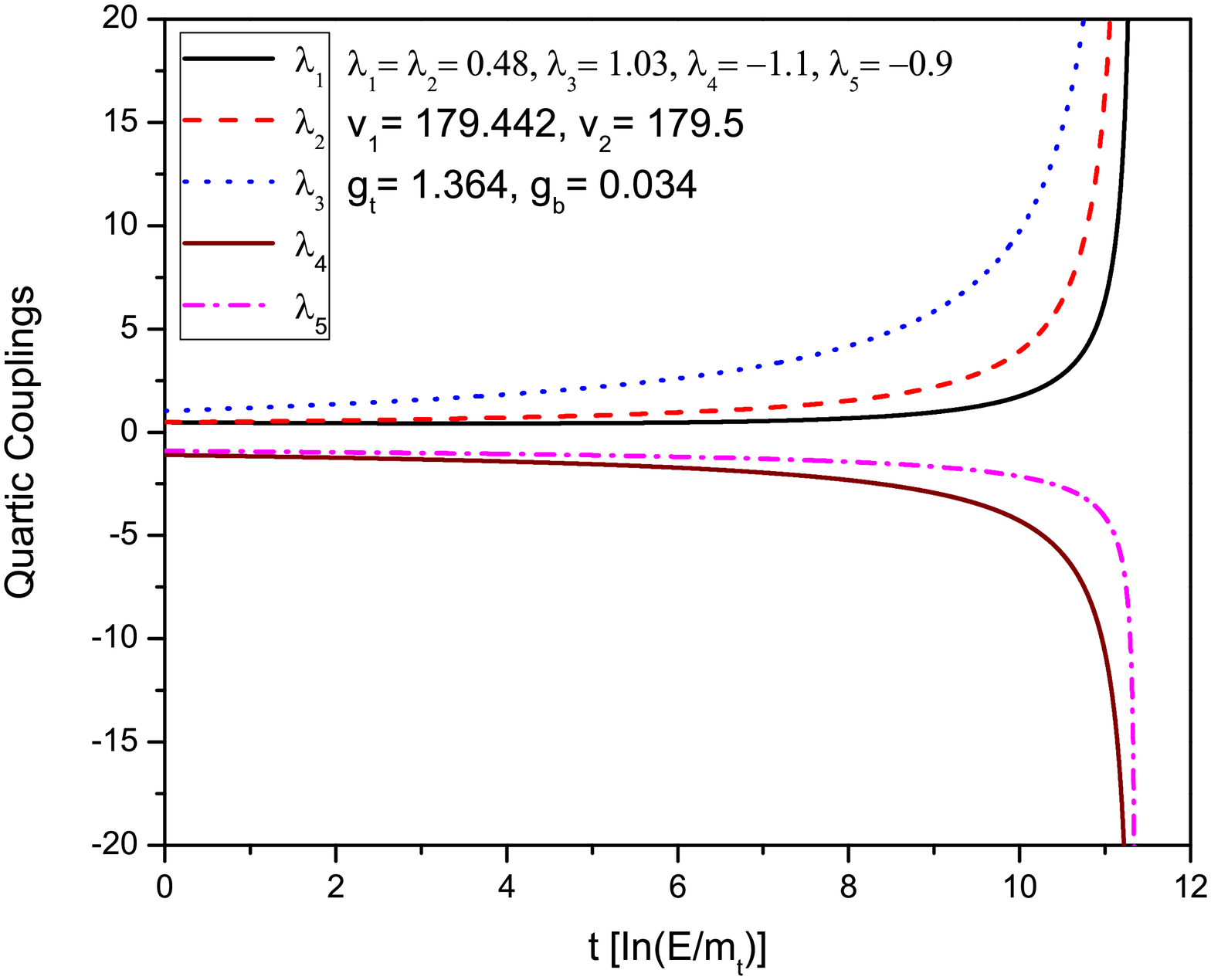}\hspace*{20pt}
\includegraphics[width=.4\linewidth]{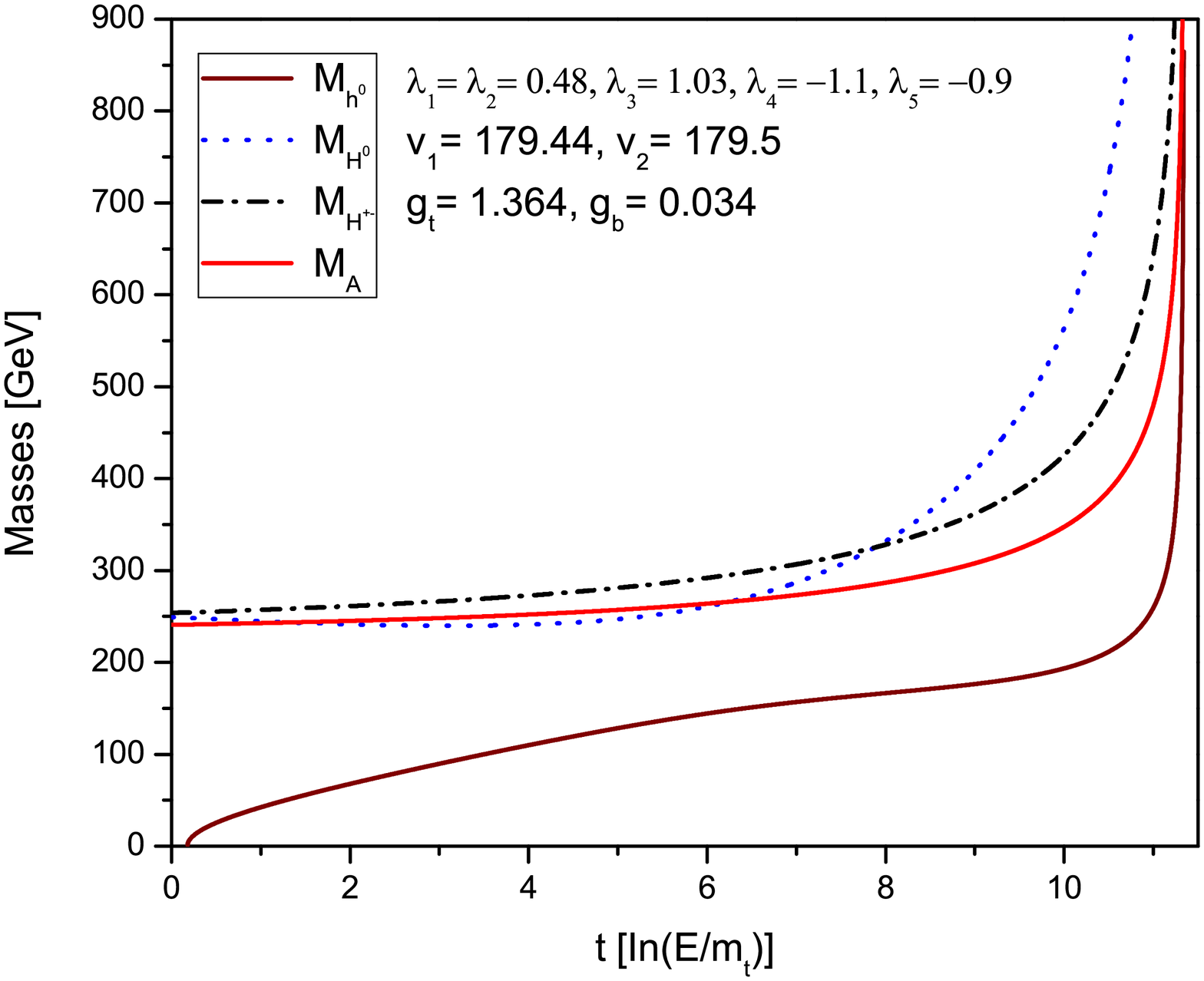}
\caption{The energy dependence of the quartic couplings and the Higgs masses,
case 1A with $\tan \beta =1$ (2D) case.\label{fig:7}}
\end{figure}

\begin{figure}[ht]
\centering
\includegraphics[width=.4\linewidth]{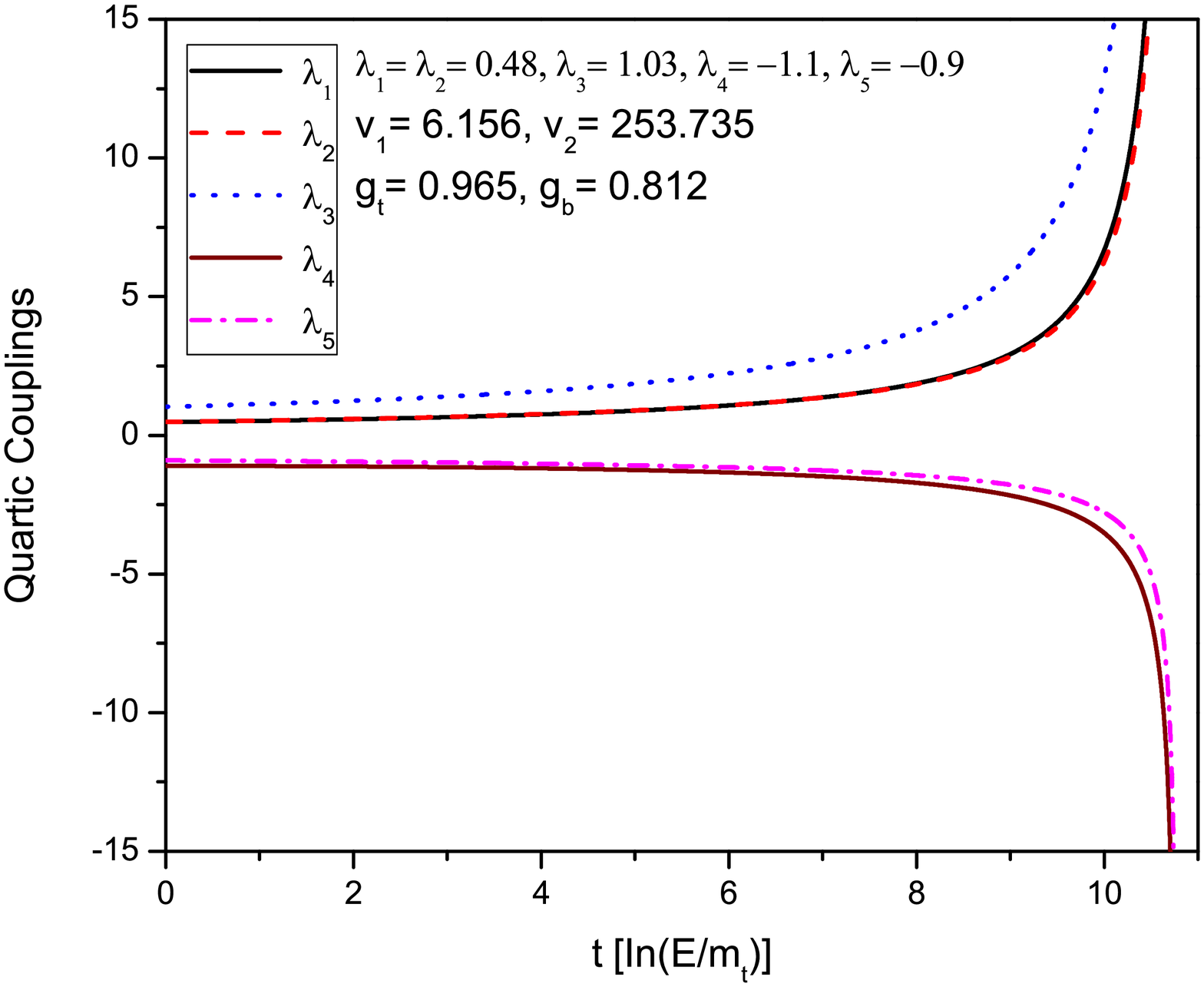}\hspace*{20pt}
\includegraphics[width=.4\linewidth]{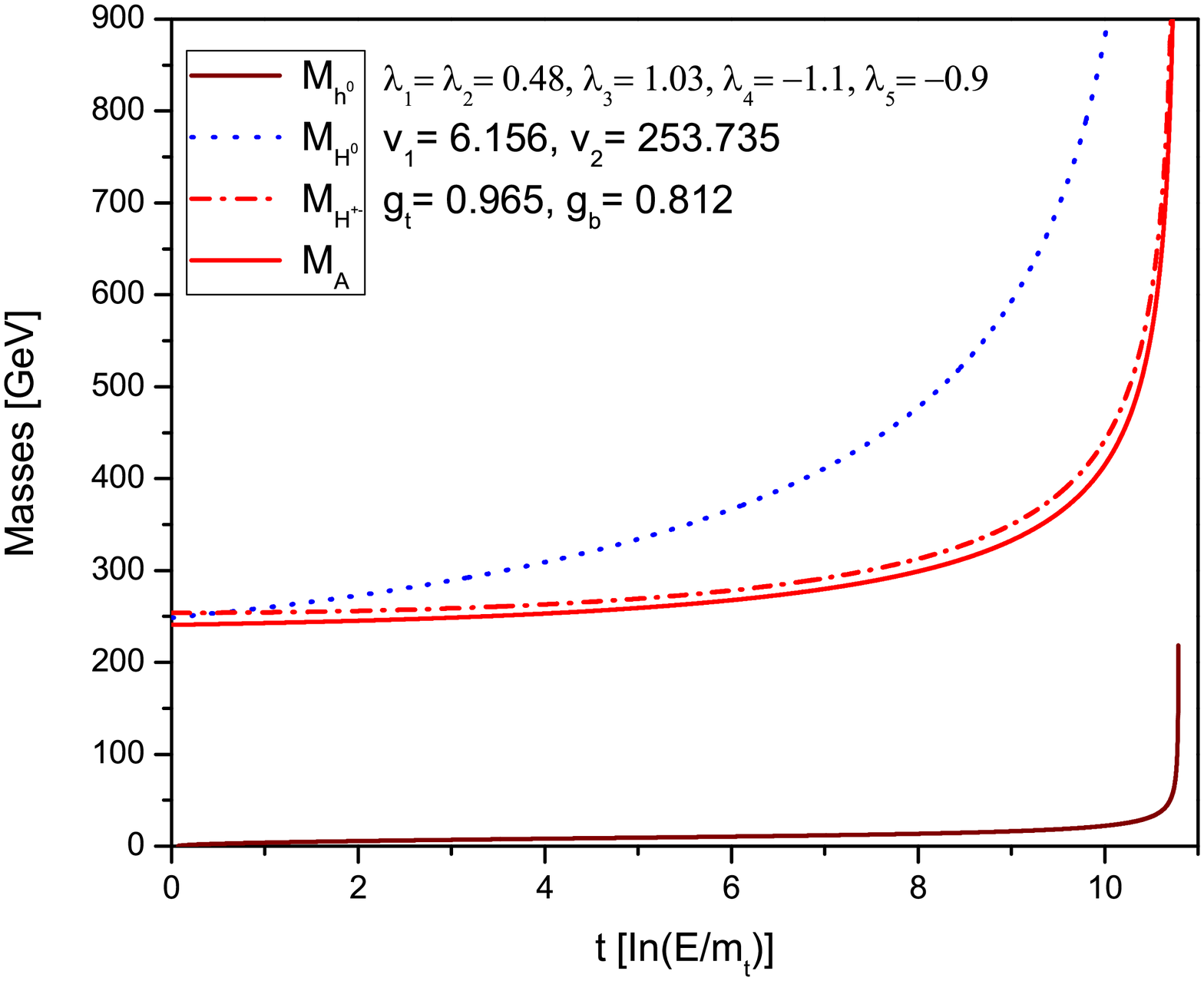}
\caption{The energy dependence of the quartic couplings and the Higgs masses,
case 1D with $\tan \beta =41.2$ (2D) case.\label{fig:8}}
\end{figure}

\begin{figure}[ht]
\centering
\includegraphics[width=0.4\linewidth]{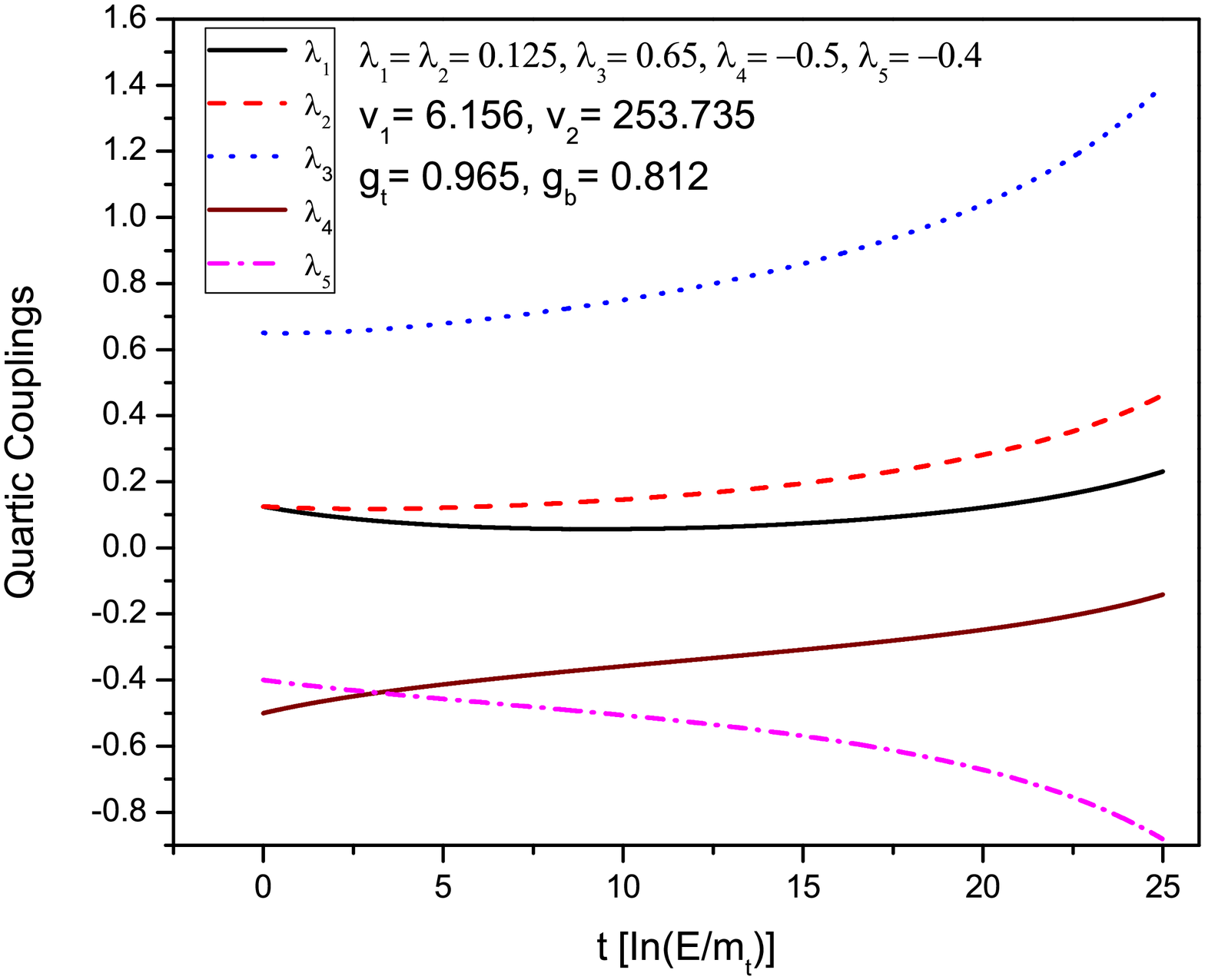}\hspace*{20pt}
\includegraphics[width=0.4\linewidth]{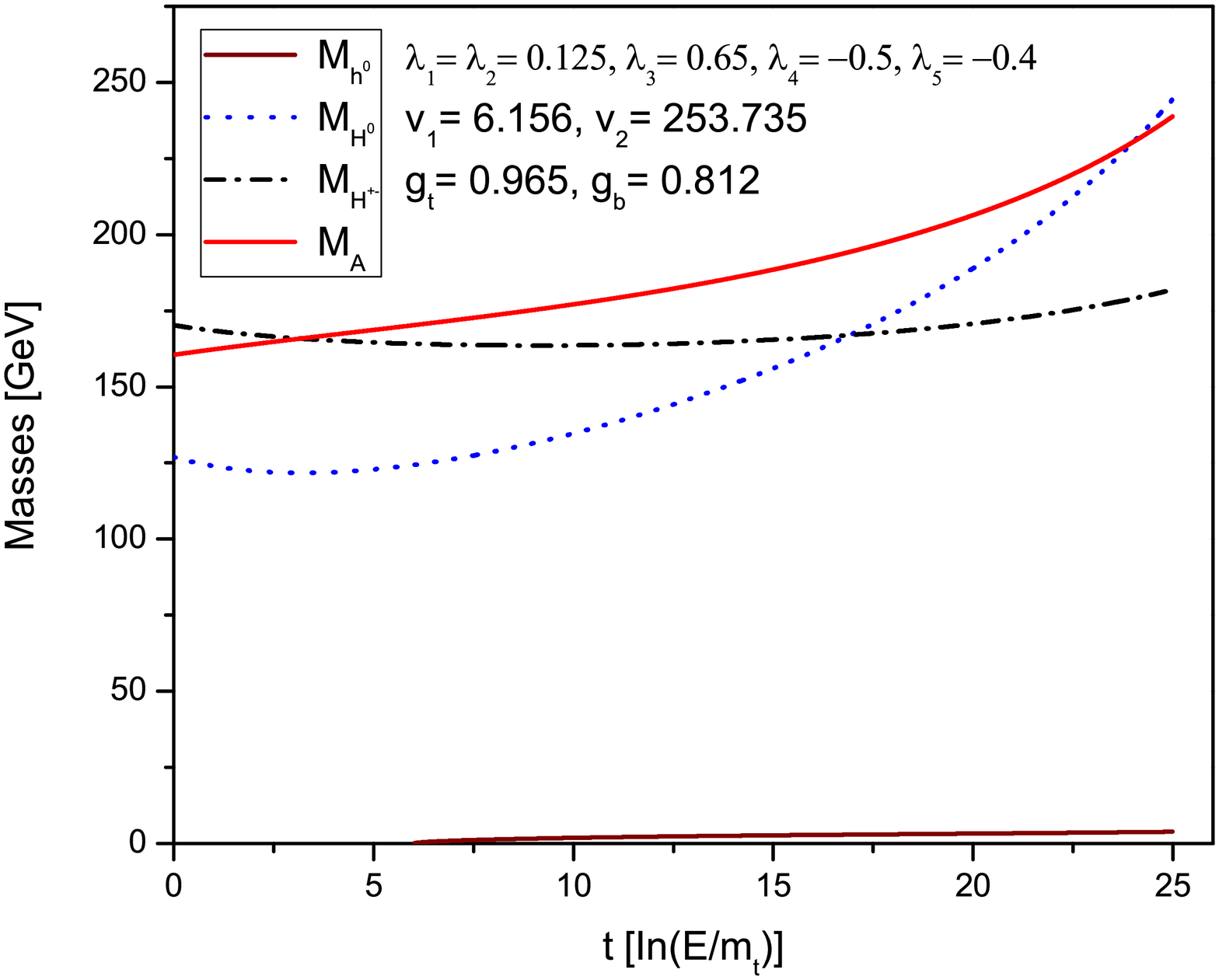}
\caption{The energy dependence of the quartic couplings and the Higgs masses,
case 1B1 with $\tan \beta =41.2$.\label{fig:9}}
\end{figure}

\begin{figure}[ht]
\centering
\includegraphics[width=.31\linewidth]{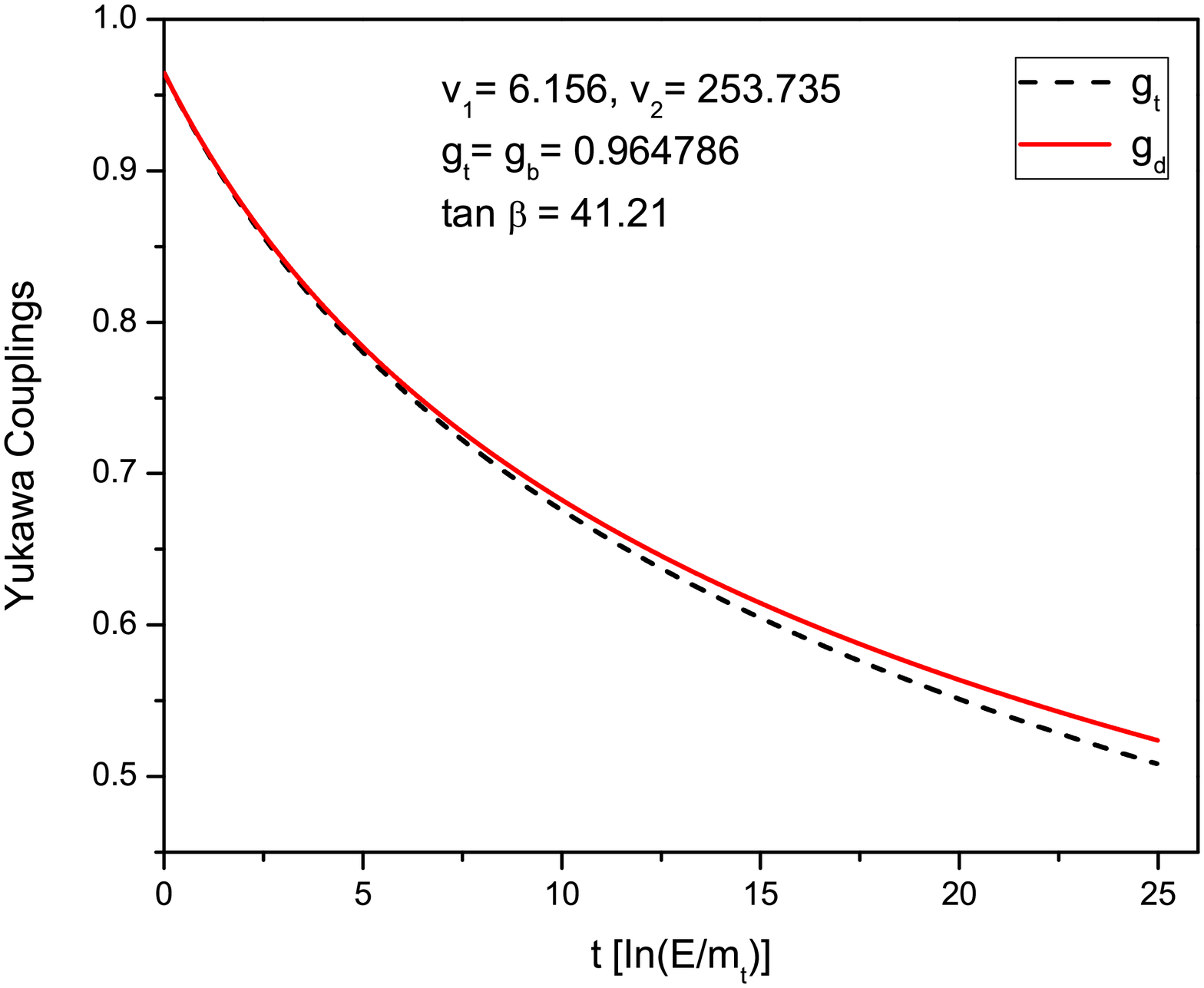}\hspace*{15pt}
\includegraphics[width=.31\linewidth]{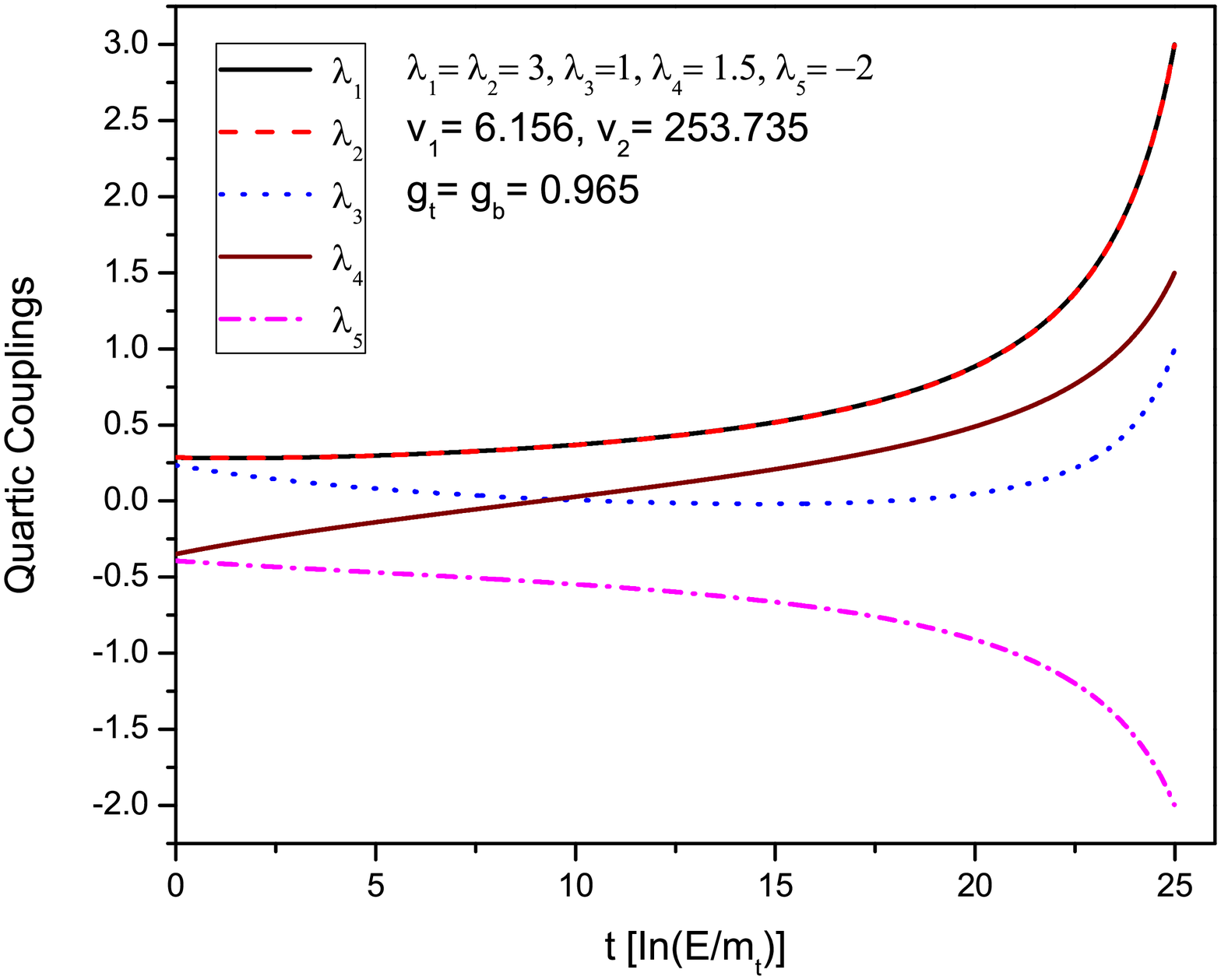}\hspace*{15pt}
\includegraphics[width=.31\linewidth]{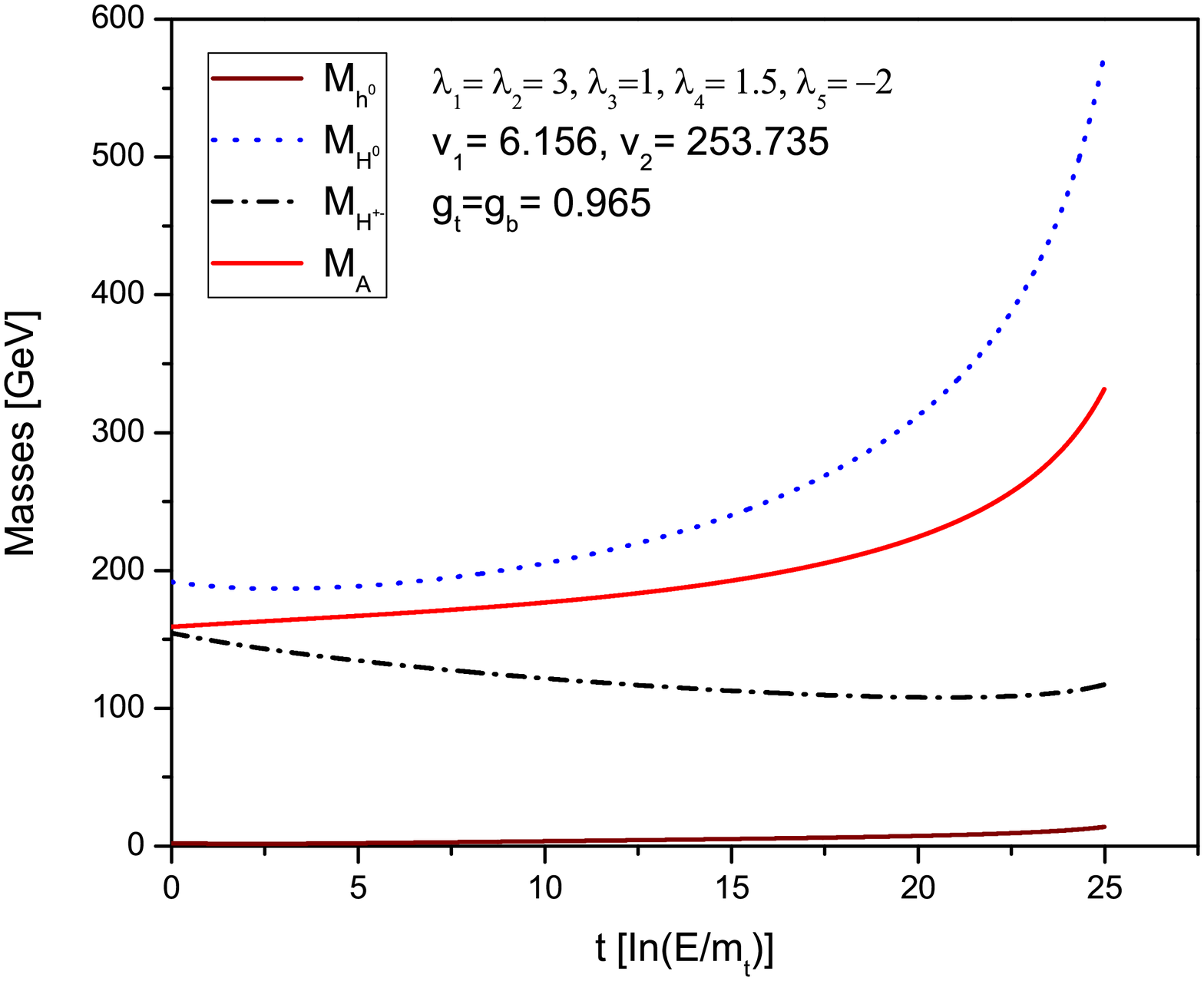}
\caption{The energy dependence of the Yukawa couplings, quartic
  couplings and the Higgs masses in the $\tan \beta =41.2$ (1D) case
  with Yukawa couplings $g_{t}=g_{b}$ at low energy.\label{fig:10}}
\end{figure}

\begin{figure}[ht]
\centering
\includegraphics[width=.31\linewidth]{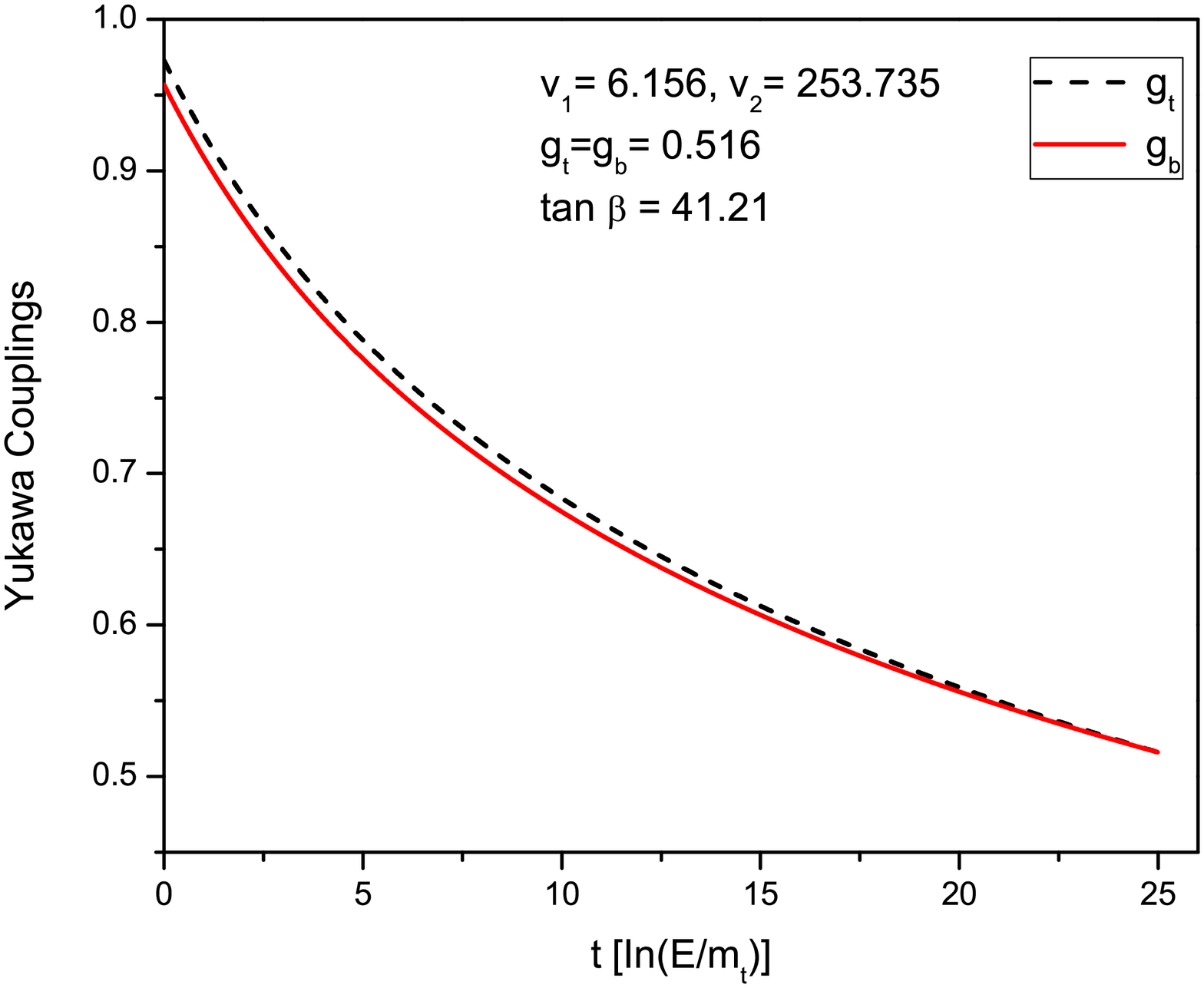}\hspace*{15pt}
\includegraphics[width=.31\linewidth]{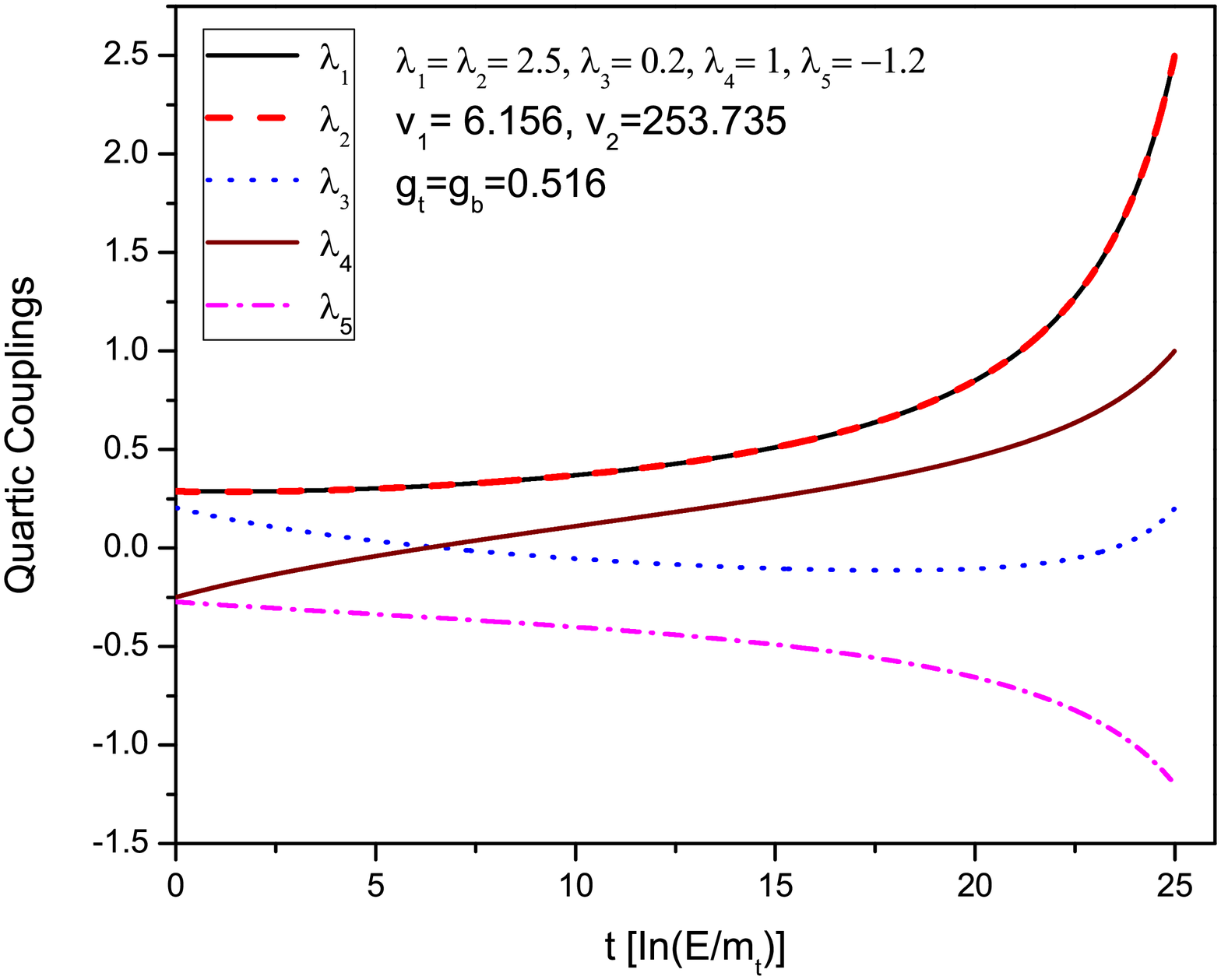}\hspace*{15pt}
\includegraphics[width=.31\linewidth]{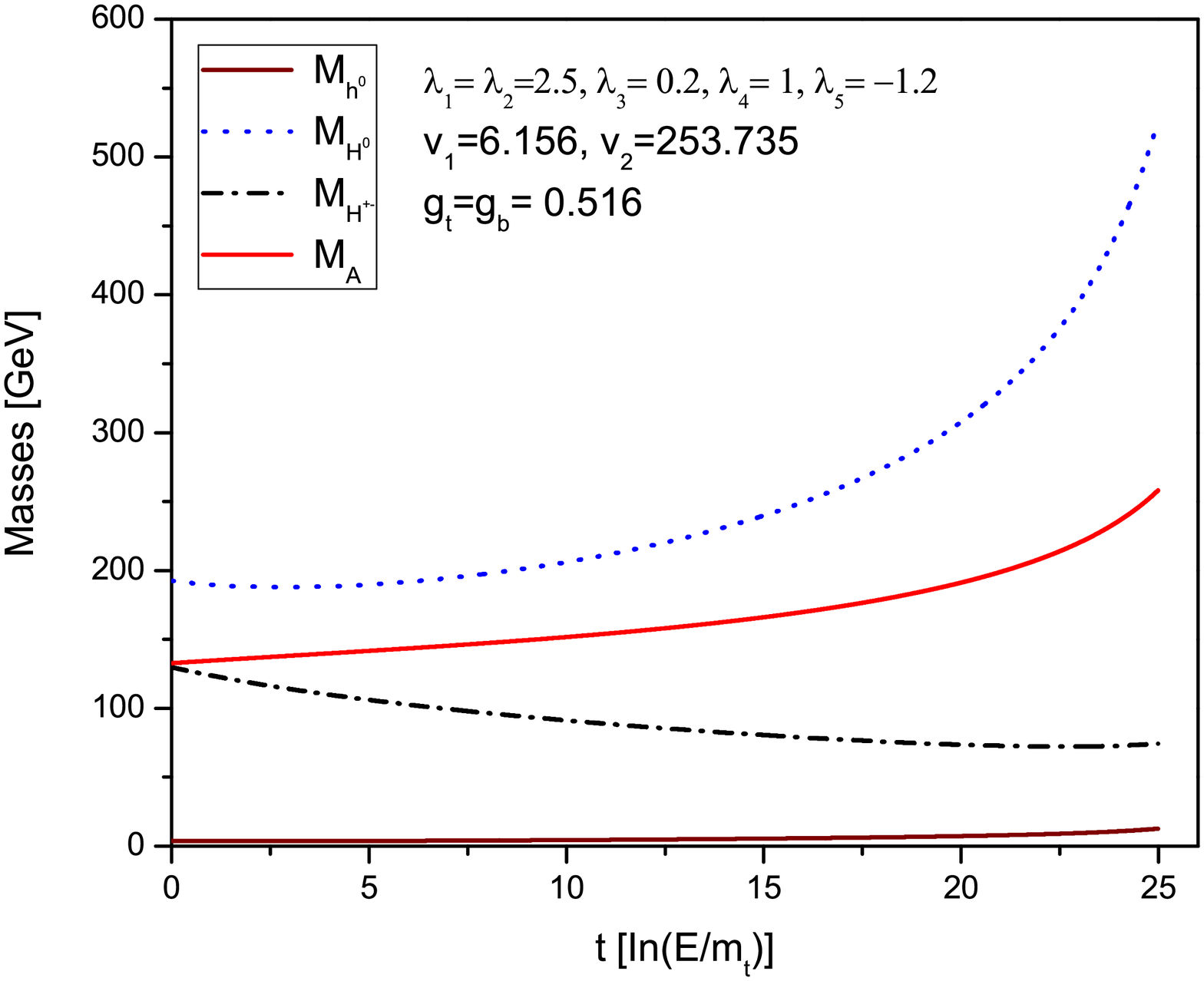}
\caption{The energy dependence of the Yukawa couplings. quartic
  couplings and the Higgs masses in the $\tan \beta =41.2$ (1D) case
  with equal Yukawa couplings $g_{t}=g_{b}$ at high
  energy.\label{fig:11}}
\end{figure}

The RGE determine the dependence of the coupling constants and other
parameters of the Lagrangian on $t$, defined as $t=\ln \left(
  E/m_{t}\right)$, where $E$ is the renormalization point energy. The
RGE for the gauge couplings $g_1$, $g_2$, $g_3$ are:
 \begin{equation}
\frac{dg_{k}}{dt}=\frac{1}{(4\pi )^{2}}b_{k}g_{k}^{3}\quad(i=1,2,3),  \label{P1}
\end{equation} 
The RGE for the Yukawa couplings of the top and bottom quarks $g_{t}, g_{b}$ are
 \begin{gather}
\frac{dg_{t}}{dt}=\frac{1}{(4\pi )^{2}}\left( \frac{9}{2}g_{t}^{2}+\frac{1}{2%
}g_{b}^{2}-(\frac{17}{20}g_{1}^{2}+\frac{9}{4}g_{2}^{2}+8g_{3}^{2})\right)
g_{t},\\
 \frac{dg_{b}}{dt}=\frac{1}{(4\pi )^{2}}\left( \frac{9}{2}g_{b}^{2}+\frac{1}{2%
}g_{t}^{2}-(\frac{1}{4}g_{1}^{2}+\frac{9}{4}g_{2}^{2}+8g_{3}^{2})\right)
g_{b}
\end{gather}
and for the vacuum expectation values $v_{1}$ and  $v_{2}$
 \begin{gather}
\frac{d}{dt}v_{1}=\frac{1}{(4\pi )^{2}}\left[ -3g_{t}^{2}+(\left(
9/20\right) g_{1}^{2}+\left( 9/4\right) g_{2}^{2})\right] v_{1},\\
\frac{d}{dt}v_{2}=\frac{1}{(4\pi )^{2}}\left[ -3g_{b}^{2}+(\left(
9/20\right) g_{1}^{2}+\left( 9/4\right) g_{2}^{2})\right] v_{2}
\end{gather}
In the equations for the quartic couplings we include the quark Yukawa
contributions of both sectors.
\begin{align}
\frac{d\lambda _{1}}{dt} =&\frac{1}{16\pi ^{2}}\left\{ 24\left( \lambda
_{1}\right) ^{2}-3\lambda _{1}\left[ 3g^{2}+\left( g^{\prime }\right)
^{2}-4g_{t}^{2}\right] +2\left( \lambda _{3}\right) ^{2}+\left( \lambda
_{4}\right) ^{2}+\left( \lambda _{5}\right) ^{2}\right.
+\left. 2\lambda _{3}\lambda _{4}+\frac{3}{8}\left( g^{2}+\left( g^{\prime
}\right) ^{2}\right) ^{2}+\frac{3}{4}g^{4}-6g_{t}^{4}\right\} ,\nonumber\\
\frac{d\lambda _{2}}{dt} =&\frac{1}{16\pi ^{2}}\left\{ 24\left( \lambda
_{2}\right) ^{2}-3\lambda _{2}\left[ 3g^{2}+\left( g^{\prime }\right)
^{2}-4g_{b}^{2}\right] +2\left( \lambda _{3}\right) ^{2}+\left( \lambda
_{4}\right) ^{2}+\left( \lambda _{5}\right) ^{2}\right. 
+\left. 2\lambda _{3}\lambda _{4}+\frac{3}{8}\left[ \left( g^{\prime
}\right) ^{2}+g^{2}\right] ^{2}+\frac{3}{4}g^{4}-6g_{b}^{4}\right\} ,\nonumber\\
\frac{d\lambda _{3}}{dt} =&\frac{1}{16\pi ^{2}}\left\{ 4\left( \lambda
_{3}\right) ^{2}+4\left( 3\lambda _{3}+\lambda _{4}\right) \left( \lambda
_{1}+\lambda _{2}\right) -3\lambda _{3}\left[ 3g^{2}+\left( g^{\prime
}\right) ^{2}-2\left( g_{t}^{2}+g_{b}^{2}\right) \right] \right.  \\
&+\left. 2\left( \lambda _{4}\right) ^{2}+2\left( \lambda _{5}\right) ^{2}+%
\frac{3}{4}\left[ g^{2}-\left( g^{\prime }\right) ^{2}\right] ^{2}+\frac{3}{2%
}g^{4}-12g_{t}^{2}g_{b}^{2}\right\} ,\nonumber\\
\frac{d\lambda _{4}}{dt} =&\frac{1}{16\pi ^{2}}\left\{ 4\left( \lambda
_{4}\right) ^{2}+4\lambda _{4}\left( \lambda _{1}+\lambda _{2}+2\lambda
_{3}\right) -3\lambda _{4}\left[ 3g^{2}+\left( g^{\prime }\right)
^{2}-2\left( g_{t}^{2}+g_{b}^{2}\right) \right] \right. 
+\left. 8\left( \lambda _{5}\right) ^{2}+3g^{2}\left( g^{\prime }\right)
^{2}+12g_{t}^{2}g_{b}^{2}\right\} ,\nonumber\\
\frac{d\lambda _{5}}{dt}=&\frac{1}{16\pi ^{2}}\lambda _{5}\left\{ 4\left(
\lambda _{1}+\lambda _{2}+2\lambda _{3}+3\lambda _{4}\right) -3\left[
3g^{2}+\left( g^{\prime }\right) ^{2}-2\left( g_{t}^{2}+g_{b}^{2}\right)
\right] \right\} .\nonumber
\end{align}
The former equations are the coupled, non linear, ordinary
differential equations whose solution provides the information about
the renormalization point energy dependence of the masses of the five
Higgs particles of the 2HDM.  To numerically solve the RGE, the
initial or final conditions for the parameters have to be previously
chosen. In order to do so we use Ref.~\cite{PDG}. The range of values,
we take, for the energy and the variable $t$ are $\left(
  E_{0}=M_{t},E_{u}\right) =\left( 173.2,1.234\cdot 10^{13}\right) $,
$\left( t_{0}=0,t_{u}=25\right)$ respectively, where $M_{t}$ stands
for the mass of the quark top and $E_{u}$ corresponds to the
electroweak unification energy where $g_{1}(E_{t})=g_{2}(E_{t})$. The
gauge couplings $%
\left( g_{1},g_{2,}g_{3}\right) _{E=M_{t}}\simeq \left(
  0.4627,0.6466,1.2367,\right) $ are obtained using the following
relations
\begin{align*}
\alpha _{e}(M_{t}) &=g_{e}^{2}/4\pi =1/\left( 127.9\right)
,\quad g_{1}(M_{t})=\sqrt{5/3}g_{e}/\cos \theta _{W}\,\quad 
\\
g_{2}(M_{t}) &=g_{e}/\sin \theta _{W},\quad g_{3}(M_{t})=\sqrt{4\pi
\alpha _{s}(M_{t})}
\end{align*}
where $\theta _{W}$ is the Weinberg angle and $\sin ^{2}\theta
_{W}(M_{t})=0.235$ and $\alpha _{s}=0.1217$.
The vev standard value that arises from 
\[
v=2M_{z}/\sqrt{g_{2}^{2}+g_{e}^{2}}
\]
is $v(M_{t})=253.81$ GeV at $M_{z}=91.19$ GeV.\\
In order to specify more rigorously the energy limits for the quartic
couplings, we have numerically solved the RGE for the gauge group
couplings $g_1$, $g_2$, $g_3$, (Fig.~\ref{fig:2}), the vacuum
expectation values $v_1$, $v_2$, and the top and the down quark Yukawa
couplings $g_t$ and $g_d$, under the following assumptions:
\begin{itemize}
\item The heaviest quark masses are related with the vevs $v_{1}$ and
  $v_{2}$ and the Yukawa couplings $g^{(u)}$ and $g^{(d)}$
\begin{equation}
  M_{t}=\frac{v_{2}}{\sqrt{2}}g_t,\quad\tan \beta=
  \frac{v_{2}}{v_{1}},\quad M_{b}=\frac{v_{1}}{\sqrt{2}}g_b.
\end{equation}
\item The gauge bosons masses are related with the gauge couplings
  $g^{\prime } $ and $g$
\begin{equation}
  M_{W}=\frac{1}{2}vg,\quad M_{Z}=\frac{M_{W}}{\cos \theta
    _{W}}=\frac{1}{2}v\sqrt{g^{2}+\left( g^{\prime }\right) ^{2}},
\end{equation}
where $\theta _{W}$ is the Weinberg angle and $e$ the electron charge
\begin{equation}
  e=g\sin \theta _{W}=g^{\prime }\cos \theta _{W}.
\end{equation}
\item Unification of the Yukawa couplings at $E=M_{t}$ or at $E_{u}$,
  i.e.,  $g_b=g_t$, and $\tan \beta = M_{t}/M_{b}$.
\end{itemize}			
It is interesting to explore now, the energy bounds of the 2DHM,
through the running of the quartic couplings which determine the mass
values of the Higgses. In the case~(c) considered in the previous
section, when $M_{H^{\pm }}=609,$ $M_{A}=621.7$, the range of validity
of the model is very short $M_{t}<E<292$ i.e., $0<t<0.52$ as can be
seen in Figs.~\ref{fig:3}--\ref{fig:6}.  There is an intermediate
class of the models depicted at Figs.~\ref{fig:7},~\ref{fig:8}, which
have an intermadite range of validity $0<t\lesssim11$. So we will
rather focus our atention on the cases where we can explore the
universality of the Yukawa couplings and its unification, to study the
mass-hierarchy problem. In this case, as can be seen in
Figs.~\ref{fig:9}--\ref{fig:11}, the 2HDM is valid in the whole range
of energies, this means $M_{t}<E< E_u$ where $E_u$ is the electroweak
unification energy. In Fig.~\ref{fig:9} we observe very slow
dependence of the quaric couplings and the Higgs masses on the
renormalization point energy. This model is characterized by rather
small values of the quartic Yukawa couplings and the value
of~$\tan\beta$ such that it permits the unification of the Yukawa
couplings of the up and down quarks $g_{t}=g_{b}$.
In Figs.~\ref{fig:10},~\ref{fig:11} we show the evolution of the
Yukawa couplings, quartic couplings and the Higgs masses for the case
when the Yukawa couplings are unified. In Fig.~\ref{fig:10} we assume
that they are unified at low energy and in Fig.~\ref{fig:11} they are
unified at high energy. The evolution of the quartic couplings and
Higgs masses are similar in both cases.
\FloatBarrier

\section{ Results and conclusions\label{sec7}}
With the aim to explore the Higgs mass content of the 2HDM extension
of the standard model, among the different forms of the Lagrangian
describing the same physical reality, we have chosen a specific one,
in which the vacuum expectation values of both Higgs fields are real,
and for simplicity also preserving the CP symmetry.  We have deduced,
in this model, the analytical expressions for the masses of the five
predicted physical Higgs particles, and expressed the Higgs potential
in terms of those masses, using Eqs.~(\ref{32}) and~(\ref{32b}).  We
have also obtained, through the mass formulas, a set of constraints to
be satisfied by the scalar parameters that determine the couplings and
self-couplings of the Higgs fields introduced in the potential
Eq.~(\ref{ec1}), and through the vacuum stability principle plus the
Lagrange Multipliers method, and obtained additional conditions to be
satisfied by those couplings.
\begin{equation}
  \lambda _{1}>0,\;\lambda _{2}>0,
\quad 4\lambda _{1}\lambda _{2}>\left( \lambda _{3}+\lambda _{4}+\lambda
_{5}\right) ^{2},\quad \left( \lambda _{4}+\lambda _{5}\right) <0, \quad \lambda
  _{5}<0,\quad \lambda _{4}<\left| \lambda _{5}\right| ,
\end{equation}
and
\begin{equation}
\lambda _{1}+\lambda _{2}>\lambda _{3}+\lambda _{4}+\lambda _{5},\quad
\lambda _{3}+\lambda _{4}+\lambda _{5}+2\sqrt{\lambda _{1}\lambda _{2}}>0,\quad
\lambda _{3}+2\sqrt{\lambda _{1}\lambda _{2}}>0,\quad\
4\left( \lambda _{1}\lambda _{2}\right) \neq \lambda _{3}^{2}.
\end{equation}
We have also looked upon extreme and semiextreme conditions on the
Higgs potential and gave a clasification of the different cases we
analized under the RGE.

As many authors base their calulations in symmetry conditions, such as
$\lambda _{1}=\lambda _{2}$ and others in a phenomenologycal study of
special events, it is important to analize the consequences of such
assumptions and we tried at least partially address this problem. 

There is a batch of data to be analysed right now in search of some of
the favored mass region, and all of it should be examined in the near
future. The results of this paper may shed some light on physics of
the Higgs sector depending on the properties of the Higgs particle.

We have considered here, symmetries in the $\lambda _{i}$ parameters,
universality of the Yukawa couplings at low energy $E_{0}$ ($M_{t}$
scale) or high energy $E_{u}\,$(weak-unification scale), hierarchy of
the quark masses and the energy range of validity of the model. The
other symmetry considered here is the unification of the Yukawa
couplings. It seems this symmetry makes the Higgs sector very stable
as can be seen in Fig.~\ref{fig:9}.

In summary, the results in this paper may be a basis for further
investigation in relation to the behavior and energy dependent
characteristics of the Higgs particles.

We finish by allegorically saying, that our paper still contains a
``blank page'', which can only be filled after the discovery of Higgs
bosons.

\end{document}